\documentclass[fleqn,usenatbib,twocolumn]{aastex631}
\shorttitle{Type Ia SNe in massive ellipticals}
\shortauthors{Mohapatra et al.}

\usepackage{graphicx}	% Including figure files
\usepackage{amsmath}	% Advanced maths commands
\usepackage{ragged2e}
\usepackage{comment}
\usepackage{capt-of}
\usepackage{bigints} %To be able to write bigger integral symbols
\usepackage{physics}
\usepackage{xcolor}
\usepackage{cleveref}
\usepackage{longtable}
\usepackage{xtab}
\usepackage{bm}
\usepackage{cancel}
\usepackage{comment}
\usepackage{CJK}

\allowdisplaybreaks
\usepackage{scalerel,tikz}

\def\mean#1{\left< #1 \right>}
\def\sun{\odot}

\defcitealias{Mohapatra2022characterising}{M22b}
\defcitealias{Mohapatra2022VSF}{M22a}
\defcitealias{Mohapatra2023MNRAS}{M23}
\defcitealias{Mohapatra2024ApJ}{MQ24}
\defcitealias{MLi2020ApJa}{MLi20a}
\defcitealias{MLi2020ApJb}{MLi20b}
\defcitealias{Voit2020ApJ}{VB20}
\begin{document}
\begin{CJK*}{UTF8}{gbsn}
%\title[SNIa and AGB in massive ellipticals]{The Supernova and AGB-Regulated Interstellar Medium: Simulations of Massive Galaxy Evolution}
\title[SNIa and AGB in massive ellipticals]{The Type Ia Supernova and Asymptotic Giant Branch Stellar Ejecta-regulated Interstellar Medium of Massive Galaxies}
% \title[SNIa and AGB in massive ellipticals]{The Type Ia Supernova and AGB-Regulated Interstellar Medium of Massive Galaxies}
% \title[SNIa in a massive ellipticals]{Unveiling the spark: How resolving Type 1a Supernovae in simulations reveals the true nature of ISM heating in massive ellipticals}
% \title[SNIa in a massive ellipticals]{The role of Type Ia supernovae in massive ellipticals: kinematics, heating, and accretion}

% The list of authors, and the short list which is used in the headers.
% If you need two or more lines of authors, add an extra line using \newauthor
% \author[Mohapatra et al]{
% Rajsekhar Mohapatra$^{\orcidicon{0000-0002-1600-7552}\,1}$\thanks{E-mail: rmohapatra@princeton.edu (RM)},
% Eliot Quataert$^{\orcidicon{0000-0001-9185-5044}\,1}$\thanks{E-mail: quataert@princeton.edu (EQ)}
% \\
\correspondingauthor{Rajsekhar Mohapatra}
\email{rmohapatra@princeton.edu}

\author[0000-0002-1600-7552]{Rajsekhar Mohapatra}
\affiliation{Department of Astrophysical Sciences, Princeton University, Princeton, NJ 08544, USA}
\author[0000-0001-9185-5044]{Eliot Quataert}
\affiliation{Department of Astrophysical Sciences, Princeton University, Princeton, NJ 08544, USA}

\author[0000-0003-3806-8548]{Drummond Fielding}
\affiliation{Department of Astronomy, Cornell University, Ithaca, NY 14853, USA}

\author[0000-0002-3680-5420]{Minghao Guo (郭明浩)}
\affiliation{Department of Astrophysical Sciences, Princeton University, Princeton, NJ 08544, USA}
% Abstract of the paper
\begin{abstract}
Observations and theory suggest that Type Ia supernovae (SNIa) heating and mass loss from asymptotic giant branch (AGB) stars play a crucial role in the interstellar medium (ISM) of massive galaxies. We perform 3D hydrodynamic simulations of the central few kiloparsecs of massive galaxies, including radiative cooling and mass and energy injection from AGB winds and SNIa (resolving each SNIa remnant, a few $\times10~\mathrm{pc}$ in size), excluding black hole feedback. We study systems with different initial core thermodynamic profiles, focusing on NGC 1399. Our simulations reproduce its observed density and entropy profiles well. Over $100~\mathrm{Myr}$, two steady-state profiles emerge, depending on the inner circumgalactic medium (CGM) pressure and the ratio of Ia heating to cooling: (i) if SNIa heating is less than cooling, a cooling flow develops; (ii) if SNIa heating is comparable to or exceeds cooling, SNIa heating drives a slow subsonic outflow of AGB ejecta, with black hole accretion at small radii. This outflow, pressure-confined by the CGM, adapts the ISM to the CGM properties: a low entropy CGM results in a dense, low entropy ISM with higher black hole accretion, while a high entropy CGM leads to a less dense, high entropy ISM with lower accretion. This suggests that the AGB-SNIa regulated ISM connects CGM and galaxy scales, potentially influencing black hole feedback in massive halos. Approximate methods of modeling Ia heating, such as clustered SNIa and smoothly distributed heating, produce unrealistic ISM profiles over $100~\mathrm{Myr}$, highlighting the importance of resolving SNIa in simulations.
\end{abstract}

%%%%%%%%%%%%%%%%%%%%%%%%%%%%%%%%%%%%%%%%%%%%%%%%%%

%%%%%%%%%%%%%%%%% BODY OF PAPER %%%%%%%%%%%%%%%%%%

\section{Introduction}\label{sec:introduction}
Massive elliptical galaxies typically lack significant star-formation, despite harboring copious amounts of hot ($T\sim10^7~\mathrm{K}$) X-ray emitting plasma in their interstellar media (ISM). In the inner few-10s of $\mathrm{kpc}$, this hot ISM can cool radiatively within less than a Hubble time and in principle supply star-forming material to the central galaxy through a cooling flow \citep{fabian1994cooling}. In addition, the old stars on the asymptotic giant branch (AGB) also supply mass to the ISM as they eject their envelopes. Hence to explain the quiescent properties of elliptical galaxies, heating mechanisms are required to balance the radiative cooling of the ISM, as well as to sweep out the excess mass deposited by the AGB ejecta \citep{Mathews1986ApJ,Ciotti1991ApJ}. 

% A paragraph about AGN heating
Jets from Active Galactic Nuclei (AGN) are generally invoked as one of the primary heating mechanisms in massive galaxies \citep{Bohringer2002A&A,mcnamara2007,Fabian2012ARA&A}he supermassive black holes (SMBHs) at the center of the galaxy accrete the infalling matter and eject mass, momentum and energy back into the ISM. 
Such AGN heating effectively acts as a time-delayed feedback loop that can respond to an increasing mass accretion rate by stronger heating and mechanical energy output; this can in turn suppress the excess accretion and keep the ISM in a statistical steady state. However, AGN jets are collimated \citep[see Fig.~2 in][]{Dunn2010MNRAS} and often deposit their energy at a few~$\times10~\mathrm{kpc}$ away from the central regions of their host galaxy. The exact spatial distribution of the energy supplied by AGN jets to the ambient ISM is thus still an open question.
 
\begin{figure}
		\centering
	\includegraphics[width=\columnwidth]{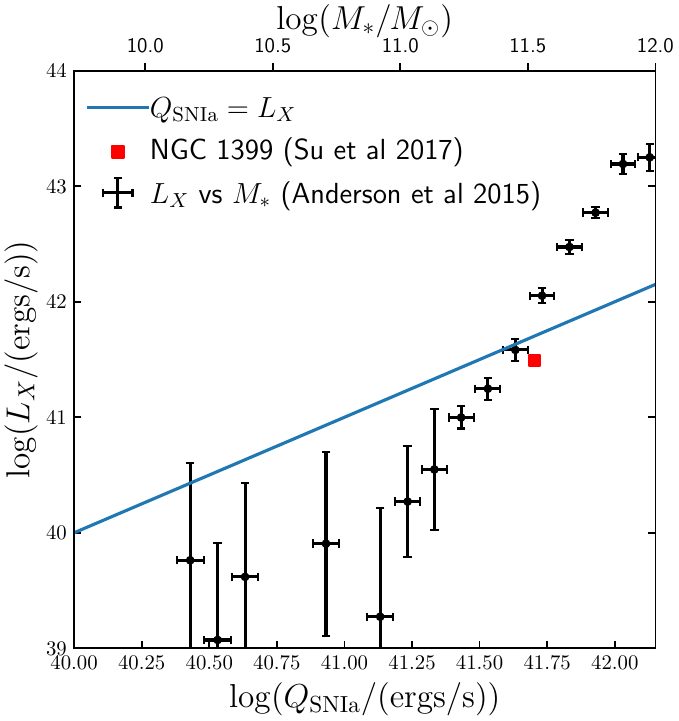}	
	\caption{X-ray luminosity versus the energy injection rate due to type Ia supernovae, assuming a $10~\mathrm{Gyr}$ old stellar population with a SNIa rate given by \Cref{eq:SNIa_rate}, and $L_X$, $M_*$ from \cite{Anderson2015MNRAS}. We import $L_X$ for NGC 1399 from \cite{Su2017ApJ} for emission within the inner $22~\mathrm{kpc}$. For NGC 1399, the net heating by the SNIa is expected to roughly balance the net X-ray emission.}
	\label{fig:MX_Mstar_Q_SNIa}
\end{figure}

In the old stellar populations of  massive galaxies, the thermalization of AGB ejecta and type Ia supernovae (SNIa) are expected to be significant channels of energy injection into the ISM. Compared to AGN jets, this stellar energy source is better understood theoretically and observationally, and more co-spatial with the hot ISM.  It is clear that there is not enough stellar heating to balance radiative cooling in the most massive galaxy groups and clusters \citep{Borgani2002MNRAS,McNamara2012NJPh,Anderson2015MNRAS}, see \Cref{fig:MX_Mstar_Q_SNIa} for reference.\footnote{Our estimate of the net SNIa heating depends on the star formation history of the galaxy, the initial mass function, etc. To estimate $Q_\mathrm{SNIa}$ in \Cref{fig:MX_Mstar_Q_SNIa}, we assume a single burst of star formation $10~\mathrm{Gyr}$ ago.}  However, in isolated elliptical galaxies and in the central parts of some groups and clusters,  heating by SNIa and mass supply by AGB stars may play a key role in suppressing star formation and determining the thermodynamic state of the ISM (e.g., \citealt{Mathews1971ApJ,Ciotti1997ApJ,Ciotti2001ApJ,Negri2014MNRAS445,Conroy2015ApJ,Generozov2015MNRAS,Voit2015ApJ803L21V}).   
More recently, using Chandra X-ray observations, \cite{Werner2012MNRAS} showed that many of the nearby massive ellipticals have similar entropy and density profiles in the central few $\sim\mathrm{kpc}$. \cite{Voit2020ApJ} (hereafter \citetalias{Voit2020ApJ}) argue that the SNIa and AGN jet heating are both important--AGN activity reduces the gas pressure at larger radii, and the SNIa heating acts in the inner regions, keeping the ISM hot and sweeping out the stellar ejecta to larger radii. \citetalias{Voit2020ApJ} showed that in many massive galaxies, the estimated heating due to SNIa and radiative cooling are in rough balance within the inner regions (see their Fig.~1). The similarity of the large-scale thermodynamic profiles across different galaxies and the near-balance between Type Ia heating and cooling in the inner regions is particularly intriguing; since the SNIa rate cannot adjust to changes in the cooling rate it is not obvious how a state with Ia heating of order cooling is realized and/or stable. If the SNIa rate exceeds the gas cooling rate, the excess energy would generate a strong wind, decreasing the gas density of the ISM and further reducing the cooling rate. Conversely, if the heating rate is lower than the cooling rate, this would result in a large-scale cooling inflow, increasing the ISM density and further enhancing its cooling rate.  \citetalias{Voit2020ApJ} argue that AGN feedback can do so because the properties of the CGM on larger scales (where AGN feedback is important) are causally connected to the galaxy scales (where AGN fueling is set and where  Type Ia SNe heating is important) by the large bounding pressure of the CGM that confines the ISM on galaxy scales.  As we shall see, our 3D simulations of the ISM of massive galaxies in the presence of Type Ia heating and AGB mass loss largely support this argument (although we do not explicitly include AGN feedback).

% How do the SNIa distribute their heat

\begin{figure*}
		\centering
	\includegraphics[width=2.0\columnwidth]{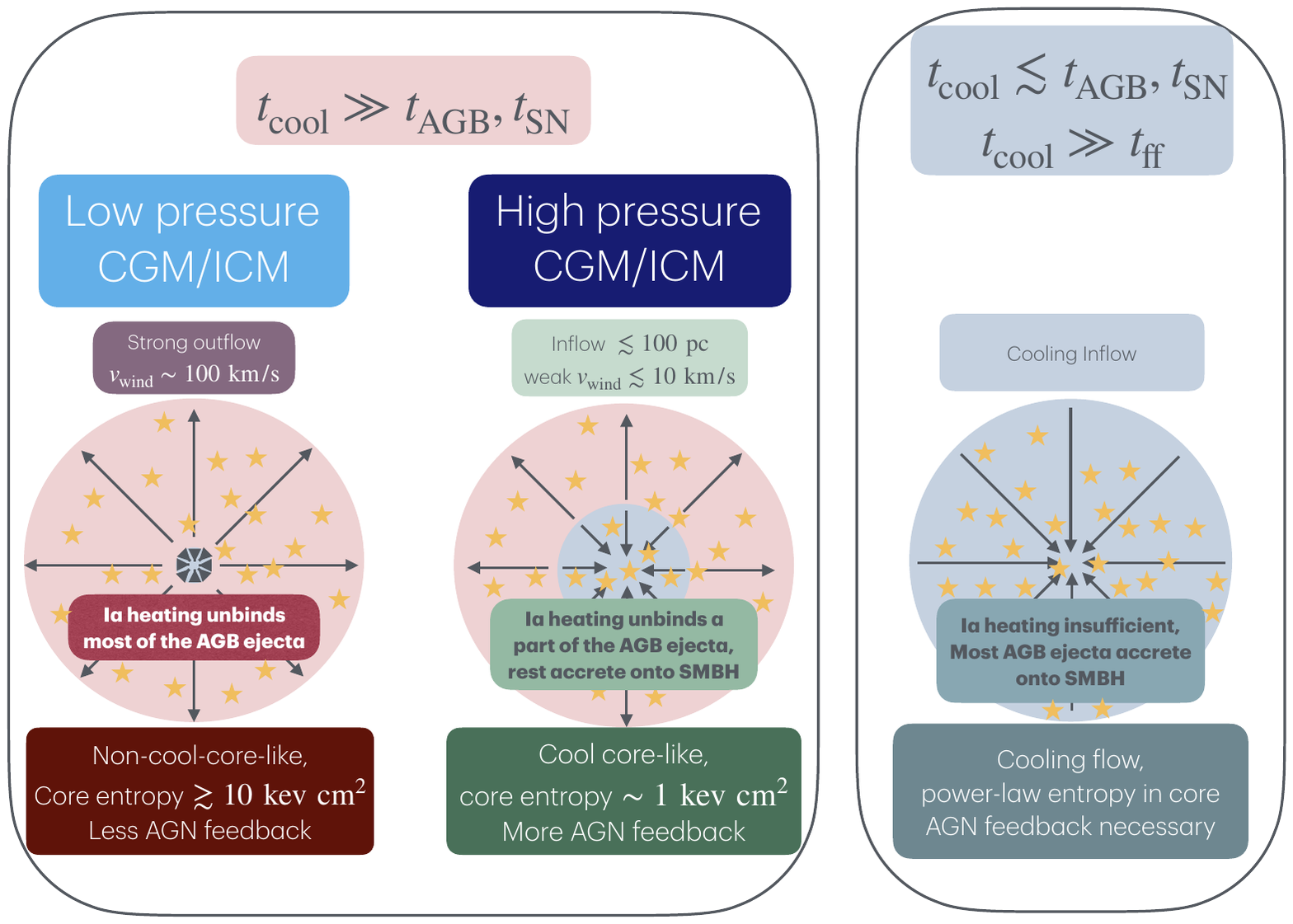}	
	\caption{A cartoon summarizing our findings on the different steady state solutions to the effects of type Ia supernovae heating and AGB mass loss on the ISM of massive ellipticals. When the cooling time is longer than the heating time, our simulations form either non-cool-core like or cool-core like entropy profiles, depending on the confining  circumgalactic medium (CGM)/ intracluster medium (ICM) pressure. When the cooling time is either comparable to, or shorter than the heating time (but longer than the dynamical time), our simulations show a cooling flow.}
	\label{fig:cartoon_diff_solutions}
\end{figure*}

Most of the models referred to above considered smooth heating of the ISM, in which the radial distribution of  heating due to the SNIa and AGB winds is assumed to follow the stellar distribution. However, \cite{YLi2019ApJ} showed that  AGB wind material could survive in the high-pressure inner ISM of  massive ellipticals without being completely thermalized. Moreover, \citet{MLi2020ApJa,MLi2020ApJb} showed using simulations of Type Ia supernovae in a homogeneous medium motivated by galaxy groups and clusters that  Type Ia heating is both inhomogeneous and efficient:   because the remnants reach pressure equilibrium with their surroundings before radiative cooling sets in, radiative losses are small but the heating is also spatially localized.   In the presence of gravity, after reaching pressure equilibrium, the Type Ia supernova remnants rise buoyantly, and deposit their energy further out from their injection sites
\citep[][hereafter \citetalias{Mohapatra2024ApJ}]{Mohapatra2024ApJ}. 

%The SNIa also drive turbulence in the ISM \citep{MLi2020ApJa,MLi2020ApJb} which affects the fraction and distribution of energy that goes into its direct thermal heating. When the SNIa heating exceeds the radiative cooling, they drive a thermal wind, although we couldn't capture global properties of the wind and the long-term evolution of the ISM due to the restrictions of our local setup. 

In this paper, we study the effects of resolved SNIa heating using hydrodynamic simulations of the central regions of an elliptical galaxy. We use observationally-motivated distributions of gas, stars and black-hole mass as our initial conditions, with a spherically symmetric gravitational potential. In addition to SNIa heating, we also account for radiative cooling, and the mass and energy deposited by an old stellar population. Our fiducial thermodynamic profiles closely match those from X-ray observations of NGC 1399. We test the effects of changing the initial gas entropy profiles, and the confining pressure of the surrounding ICM/CGM. Our results build on earlier work by \cite{Tang2009MNRASa,Tang2009MNRASb}  who  considered the effect of resolved SNIa heating in the Milky Way's bulge.  
%In comparison to a smooth stellar heating model, the resolved SNIa heating leads to non-uniform distributions of  density, temperature, and metallicity and increases the X-ray luminosity. However, 
The ISM in the Milky Way's bulge is at a much lower temperature, density and a shallower gravitational potential compared to the center of an elliptical galaxy-- it is thus non-trivial to extend their results to  massive galaxies.   In our calculations, we intentionally do not include AGN feedback since we are interested in understanding the implications of AGB mass-loss and Type Ia SNe for massive galaxy evolution.   
 
In \Cref{fig:cartoon_diff_solutions}, we present a heuristic summary of our key findings. We find that the thermodynamic properties in the inner several $\mathrm{kpc}$ of massive ellipticals are set by a combination of the mass deposited by AGB stars, energy injected due to Ia supernovae, and the confining pressure of the circumgalactic medium (CGM)/ intracluster medium (ICM). As also argued by \citetalias{Voit2020ApJ}, there are two different classes of solutions:
\begin{enumerate}
    \item The properties of the AGB ejecta heated by Type Ia supernovae on scales of the galaxy depend on the confining pressure of the surrounding CGM/ICM (we explain this analytically in \S \ref{sec:theoretical_background}).   For a low confining pressure, the SNIa drive a strong (albeit still subsonic) $\sim100~\mathrm{km/s}$ wind and keep the ISM low density, with entropy $\gtrsim10~\mathrm{keV}\mathrm{cm}^2$. We denote this solution as a `non-cool-core' like state.  When the ambient CGM/ICM has a higher pressure, the outflow driven by  SNIa is slower $\lesssim10~\mathrm{km/s}$ and denser, in a `cool-core' like state, with core entropy $\sim1~\mathrm{keV}\mathrm{cm}^2$. The ISM in this state best matches X-ray observations of many nearby massive ellipticals. These solutions are analogous to the low-pressure and high-pressure CGM in \citetalias{Voit2020ApJ} with stellar velocity dispersion $\sigma_*\gtrsim300~\mathrm{km/s}$.    
    \item We find that a second class of solutions arises when the density of the galaxy's ISM is yet higher, such that the cooling time  is either comparable to, or shorter than the heating time due to Type Ia supernovae (but longer than the dynamical time). The entire ISM then undergoes a cooling flow to smaller radii.  
\end{enumerate}

% \cite{Generozov2015MNRAS} calculated steady-state 1D hydrodynamic profiles including different heating processes using observationally motivated stellar density profiles. They derive analytic estimates for the stagnation radius and the gas accretion rate onto the central black hole. 

The rest of this paper is organized as follows:  In \S \ref{sec:theoretical_background} we provide some theoretical background that explains the types of solutions illustrated in Figure \ref{fig:cartoon_diff_solutions}. Then we describe our methods and list our suite of simulations in \S\ref{sec:Methods}. In \S\ref{sec:resolved_sn1a_feedback}, we elaborate further on our key results that we summarized above in Fig.~\ref{fig:cartoon_diff_solutions}. We compare between modeling SNIa as discrete, resolved events versus other approximate injection methods in \S\ref{sec:approx_1a_modeling}. We discuss simulations that zoom in to the region close to the black hole and zoom out to study the longer timescale evolution of the system in \S\ref{sec:zoom_in_zoom_out}.  We summarize and conclude in \S\ref{sec:Conclusion}.

%In the latter part of the paper, we study the effects of different SNIa rates, which changes the effective entropy of the injected material. We also compare between the heating due to randomly injected resolved SNIa versus other approximate methods of modeling SNIa, such as smooth heating (that only increases the internal energy of the ISM based on the stellar distribution and the expected Ia rate) and clustered SNIa injection (where each event carries more energy, but the event rate is smaller). We study their effects on the evolution of gas density and entropy profiles, the mass accretion rate, the angular momentum of accreted gas, the distribution of stellar and SNe ejecta, and the primary channels of energy transport to the ISM. We also perform perform zoom-in simulations, where we resolve the gas accretion down to the circularization radius $\sim0.01~\mathrm{pc}$, and zoom-out simulations, where we study the long-term evolution (upto $3~\mathrm{Gyr}$) of the gas thermodynamic profiles out to $r=20~\mathrm{kpc}$.

% Hence one needs to capture the evolution of resolved SNIa remnants to model the heat distribution.  
% Li et al papers ....
% Our last paper ....

% Resolved SNIa heating - local sims and Tang et al papers for the MW's Bulge
% Ciotti et al papers
% metallicity evolution?
\section{Theoretical background} \label{sec:theoretical_background}

To interpret our numerical results that follow it will be useful to briefly consider a simplified  model of some of the physics associated with the interaction of mass-loss from evolved stars and Type Ia supernovae.   Our goal in doing so is not to quantitatively reproduce the simulations, although it is likely that, e.g., an extension of the models in \citet{Generozov2015MNRAS,Voit2020ApJ} could do so.\footnote{One feature of our simulations in \S \ref{sec:resolved_sn1a_feedback} and \ref{sec:approx_1a_modeling} is that the net effect of Type Ia SNe is not fully local, i.e., not directly proportional to the local stellar density; this is because the remnants can rise buoyantly and/or drive convection.   This could potentially be accounted for analytically with an additional radial flux of energy in spherically symmetric models.}  Rather, our goal here is to semi-quantitatively motivate the regimes described in Figure \ref{fig:cartoon_diff_solutions}, which are indeed all realized in our simulations described in \S \ref{sec:Methods} and \ref{sec:resolved_sn1a_feedback}.   Many of the ideas here related to the bounding CGM pressure on Type Ia SNe driven winds in massive galaxies were nicely elucidated first in \citetalias{Voit2020ApJ}, although our reasoning is somewhat complementary to theirs (we draw more explicitly on the galactic wind and star-cluster wind analogy).

In the old stellar populations focused on in this paper, stellar evolution produces a mass-loss rate of $\dot M_\star = M_\star/4 \times 10^{11} {\rm yrs} = 0.25 (M_\star/10^{11} M_\odot) M_\odot \, {\rm yr^{-1}}$  \citep{Conroy2009ApJ}.  The stellar velocity dispersion is much larger than the wind speed $\sim 10-50 \, {\rm km \, s^{-1}}$ of the AGB winds that dominate stellar mass loss in evolved stellar populations.   We assume that the AGB wind material shocks on the surrounding ISM, thermalizes, and is incorporated into it (we use this same assumption in our simulations described in \S \ref{sec:Methods}).   The dominant heating of the ISM is provided by Type Ia SNe, which occur at a rate $\dot N_\mathrm{Ia} \simeq 0.006 \, {\rm yr^{-1}} \, (M_\star/10^{11} M_\odot)$    given the delay-time distribution from \cite{Maoz2017ApJ}.  The energy per unit mass provided by stellar evolution defines a critical velocity scale of 
\begin{equation}
    v_{crit} = \left(\frac{2 \dot N_\mathrm{Ia} E_\mathrm{Ia}}{\dot M_\star}\right)^{1/2} \simeq 1000 \, {\rm km \, s^{-1}}
    \label{eq:vcrit}
\end{equation}
where the energy per Type Ia SN is $10^{51}$ ergs.  Note that AGB gas heated by Type Ia supernovae will reach a temperature such that $\frac{2}{5} kT/m_p \simeq \frac{1}{2} v_{crit}^2$ (where the 2/5 is set by the enthalpy of an ideal gas) which corresponds to $\simeq 1.5 \times 10^7$ K or a sound speed of $\simeq 450 {\rm km \, s^{-1}}$. This is coincidentally similar to or slightly larger than the velocity dispersions of massive galaxies.   
%The parameter $v_{crit}$ in equation \ref{eq:vcrit} is analogous to $\epsilon_\star$ in (\citetalias{Voit2020ApJ}).

Neglecting for the moment gravity and radiative losses, the combined effect of AGB mass loss and Type Ia SNe would be to drive a galactic wind with properties analogous to those described in models of galactic winds driven by core-collapse supernovae \citep{Chevalier1985} or models of star cluster winds produced by the interaction of colliding stellar winds \citep{Canto2000}.  
%In contrast to these applications, the mass-loading of the wind in our case is set by AGB mass loss and the heating (energy-loading) by Type Ia SNe.  
As described in \citet{Canto2000} there are two classes of solutions depending on the pressure of the surrounding medium -- in our problem this should be interpreted as the circumgalactic medium or intracluster medium on scales of $\sim 5-10$ kpc just exterior to the galaxy where the AGB winds and Ia supernovae are primarily located.   If the bounding pressure is small the galaxy-scale wind driven by Type Ia SNe is supersonic with an asymptotic velocity $\sim v_{crit}$.   If the bounding pressure is large, however, this suppresses the supersonic wind, leading instead to a slow subsonic `breeze'.  Substituting $\gamma=5/3$ in equation 24 of \cite{Canto2000}, the critical pressure below which the supersonic wind develops is
\begin{equation}
P_{ss} \simeq \frac{0.43 \dot M_\star v_{crit}}{4 \pi R^2} \sim 4 \times 10^{-13} \, {\rm erg \, cm^{-3}}
\label{eq:Pss}
\end{equation}
where in the first expression $R$ is roughly the size of the galaxy and in the numerical evaluation we used $R \sim R_{e} \simeq 4 (M_\star/10^{11} M_\odot)^{0.56}$ kpc for massive elliptical galaxies \citep{Shen2003} and have neglected the galaxy stellar mass dependence given how weak it is ($P_{ss} \propto M_{\star}^{-0.1}$).   Equation \ref{eq:Pss} is essentially the requirement that the ram pressure of the wind at radius $R$ must exceed the ambient bounding pressure for a supersonic wind to develop.  If we take a fiducial $k T \sim$ 1.5 keV (motivated by the temperature implied by eq. \ref{eq:vcrit}) the critical pressure in equation \ref{eq:Pss} is equivalent to a density of $\sim 10^{-4}/(kT/1.5 \, {\rm keV})$ cm$^{-3}$ or an entropy of $\sim 700 \, (kT/1.5 \, {\rm keV})^{5/3}$ keV cm$^2$.   This is a quite low density (high entropy) by the standards of the inner CGM/ICM.   In practice this means that the bounding pressures of interest are $\gg P_{ss}$ and thus Type Ia supernovae mass-loaded by AGB winds cannot drive supersonic outflows.  They can at most instead drive slow roughly hydrostatic outflows whose properties depend on the bounding pressure set by the CGM/ICM $P_{CGM}$ (as in the schematic Figure \ref{fig:cartoon_diff_solutions}).  In particular, the above reasoning demonstrates that the effect of Type Ia supernovae is to regulate the ISM on galactic scales to be a hydrostatic extension of the CGM/ICM, i.e., to satisfy $dP/dr \sim -\rho g$  with the boundary condition that $n \sim n_{CGM}$ and $P \sim P_{CGM}$ at $r \sim  R_e$.  This is true even though in principle the source of mass can be completely different interior to (AGB winds) and exterior to (cosmological accretion) $\sim R_e$.   Quantitatively, if the outflow rate at $\sim  R_e$ is the total mass-loss rate supplied by stellar evolution, then the velocity of the outflow we expect is
\begin{equation}
v(R_e) \sim \frac{\dot M_\star}{4 \pi R_e^2 \rho_{CGM}} \sim 3 \, {\rm km \, s^{-1}} \, \left(\frac{n_{CGM}}{0.02 {\rm cm^{-3}}}\right)^{-1}
\label{eq:vout}
\end{equation}
where we assume that $n_{CGM}$ -- the density of the CGM at $\sim R_e$ -- is largely set by other physics and we again neglect the weak galaxy stellar mass dependence. %If the galaxy stellar density profile is $\rho \propto r^{-\alpha}$ for radii of interest then $\dot M_\star \propto r^{3-\alpha}$ and $v(r) \simeq v(2 R_e) (r/2 R_e)$
Equation \ref{eq:vout} is a highly subsonic flow for all expected ambient CGM densities/pressures.  Note also that the flow time across a region of $\sim R_e$ given equation \ref{eq:vout} is $\sim 10^{9}$ yrs so the SNe-driven wind/breeze is unlikely to sweep that far into the inner CGM.

A key feature of the physics described here is that the AGB/Ia regulated ISM on galactic scales self-adjusts to adopt to the bounding properties set by the surrounding CGM/ICM: for exactly the same central galaxy, a non-cool core ambient cluster with low density and high entropy will produce a low density and high entropy ISM on galactic scales, while a cool-core ambient cluster with higher density and lower entropy will produce a higher density and lower entropy ISM on galactic scales.   This is possible for the same central galaxy and stellar mass loss rate because large differences in density, entropy, and pressure can be accommodated for by changes in velocity (eq. \ref{eq:vout}) while the velocity is at all times subsonic and so does not significantly modify the hydrostatic or thermal structure of the ISM/CGM.  As a result, the AGB and Ia regulated ISM on galactic scales maintains the broad thermodynamic properties (high or low pressure, entropy, density) of the inner CGM even though the latter can be regulated by very different physics (e.g., AGN feedback; \citetalias{Voit2020ApJ}), as we will return to in \S \ref{sec:Conclusion}.
%This is because of the effect of the bounding pressure of the CGM on the galaxy-scale AGB ejecta and Type Ia SNe.

A key assumption in the above reasoning is that Type Ia heating is able to offset radiative cooling of the ISM on galactic scales.   Not surprisingly, this fails to be true for a sufficiently dense CGM/ICM, in which case our models here predict (absent other physics such as AGN) that a cooling flow will develop, as illustrated in the right column of Figure \ref{fig:cartoon_diff_solutions}.

\section{Methods}\label{sec:Methods}

This section describes the setup and methods for our numerical simulations of the interplay between AGB mass loss and Type Ia supernovae in massive galaxies, including our fiducial galaxy model, model equations, and numerical methods.

\subsection{Galaxy Model and Initial Conditions}\label{subsec:dens_entr_profiles}

%implementation of stellar heating and radiative cooling, and presents key arguments based on timescale analysis.
We choose NGC1399, the early-type galaxy at the center of the Fornax cluster as a representative case study for the subsequent calculations.    We construct the gravitational potential by taking into account the contribution from the central supermassive black hole ($M_\mathrm{BH}=1.06\times10^9M_\sun$), a Hernquist profile for stars in the stellar bulge with $M_\mathrm{*}=3.6\times10^{11}M_\sun$ and an effective radius $r_\mathrm{eff} = 3.6~\mathrm{kpc}$ \citep{Merritt2001MNRAS}, and an NFW profile for the dark matter distribution, with $M_\mathrm{dm}=10^{13}~\mathrm{M}_\sun$ and $r_s=50~\mathrm{kpc}$, which describes the inner dark matter profile well \citep{Richtler2008A&A}. 
% The mass enclosed within a radius $r$ is thus given by:
    
% \begin{align}
%     M(r) &= M_\mathrm{dm}(\log_{10}\left(1+r/r_s)-r/(r_s+r)\right)\label{eq:mass_enclosed}\\
%     &+ M_\mathrm{*}\left(r/(r+r_\mathrm{eff}/1.8)\right)^2 +M_\mathrm{BH},\nonumber
% \end{align}
% and the acceleration due to gravity $\bm{g}$ is given by 
% \begin{equation}
%     \bm{g} = -GM(r)/r^2 ~\hat{\bm{r}}.\label{eq:grav_acc}
% \end{equation}
% Heating vs cooling rate for NGC1399
%tcool, t_AGB, t_SN, tff at initial time?

%\subsection{Radiative cooling}\label{subsec:theory_cooling}

In our default model, we set the initial radial profile of gas entropy ($S$) at $t=0$ as: 
\begin{equation}
    S=S_0(1+r/r_S)^{\alpha_S},
\end{equation}
where $S_0=2.88~\mathrm{keV}\mathrm{cm}^{2}$, $r_S=0.5~\mathrm{kpc}$, and $\alpha_S=1.1$  (since the observed profiles in \citealt{Werner2012MNRAS} scale roughly as $r^1$ and flatten towards the center). We construct the density profile by setting
the number density $n=n_0=0.3~\mathrm{cm}^{-3}$ at $r=r_S$ and solving for hydrostatic equilibrium inward and outward from $r=r_S$, respectively \citep[similar to the method employed in][]{MGuo2023ApJ}. 
The density and entropy profiles at $t=0$ are shown in Fig.~\ref{fig:init_rad_prof_dens_entr_timescales} and are in rough agreement with the X-ray observations of \cite{Werner2012MNRAS}.   As summarized in \S \ref{subsec:list_of_simulations}, in addition to our fiducial model motivated by observations of NGC 1399, we also consider a range of different initial gas pressures and temperatures to mimic other observed systems.

% === Methods ===
%\section{Methods}\label{sec:Methods}

\subsection{Simulated Equations}\label{subsec:ModEq}
We model the ISM as an ideal gas with an adiabatic index $\gamma=5/3$. We evolve the following equations:

\begin{subequations}
	\begin{align}
	\label{eq:continuity}
	&\frac{\partial\rho}{\partial t}+\nabla\cdot (\rho \bm{v})=\dot{\rho}_\mathrm{Ia}+\dot{\rho}_*,\\
	\label{eq:momentum}
	&\frac{\partial(\rho\bm{v})}{\partial t}+\nabla\cdot (\rho\bm{v}\otimes\bm{v})+ \nabla P=\rho \bm{g} + \dot{\bm{p}}_\mathrm{Ia},\\% + (\cancel{ \dot{n}_\mathrm{Ia}M_\mathrm{Ia}}+\dot{\rho}_*)\bm{v},\\
	\label{eq:energy}
	&\frac{\partial e}{\partial t}+\nabla\cdot ((e+P)\bm{v})=-\rho (\bm{v}\cdot\nabla)\Phi \\
    &\qquad \qquad \qquad+\dot{e}_\mathrm{Ia}+\dot{e}_*-\mathcal{L}, \nonumber\\
	&e=\frac{\rho\bm{v}\cdot\bm{v}}{2} + e_\mathrm{int},\label{eq:tot_energy}\\
    &\frac{\partial (\rho C_i)}{\partial t}+\nabla \cdot(\rho \boldsymbol{v}C_i) = \dot{\rho}_i, \label{eq:scalar}
	\end{align}
    \end{subequations}
where $\rho$ is the gas mass density, $\bm{v}$ is the velocity, $P=\rho k_B T/(\mu m_p)$, where $\mu\approx0.61m_p$ is the mean particle weight, $m_p$ is the proton mass and $k_B$ is the Boltzmann constant. In the energy equation (\ref{eq:energy}), the total energy density is given by $e$ and the gravitational potential is denoted by $\Phi$, with $\bm{g}=-\nabla\Phi$. The internal energy density $e_\mathrm{int}=P/(\gamma-1)$. In equation (\ref{eq:scalar}), $C_i$ is the specific density of passive scalar species to trace stellar ejecta from AGB and SNIa, respectively. The remaining source terms due to stellar evolution and Type Ia supernovae are discussed in \S \ref{sec:theoretical_background} and are given by
%We assume the stellar population to be $10~\mathrm{Gyrs}$ old and that their specific mass loss rate is given by: 
\begin{equation}
    \dot{\rho}_*/\rho_* = 2.5\times10^{-6} \, \mathrm{Myr}^{-1},\label{eq:stellar_mass_rate}
\end{equation}
where $\rho_*$ is the stellar density %\citep[assuming evolution models of][]{Conroy2009ApJ}. Additionally, we assume that all of this material is thermalized in the ISM, which gives an energy injection rate per unit volume:
%\begin{subequations}
\begin{equation}
    \dot{e}_* = (3/2)\dot{\rho}_*\sigma_*^2,\label{eq:AGB_wind_heat_rate_estimate}
\end{equation}
where for simplicity we use a constant stellar velocity dispersion of $\sigma_*=300~\mathrm{km/s}$ in the stellar heating term (which is subdominant relative to Ia heating regardless), and
%is the stellar velocity dispersion \citep{Merritt2001MNRAS}, and
%\subsection{Heating due to SNIa}\label{subsec:theory_SN_injection}
%We expect type Ia supernovae go off at %a rate density:
\begin{equation}
    \dot{n}_\mathrm{Ia} = 600 \, \mathrm{kpc}^{-3} \, \mathrm{Myr}^{-1} \left(\frac{\rho_*}{10^{10}~M_\sun \, \mathrm{kpc}^{-3}}\right)
    \label{eq:SNIa_rate}
\end{equation}
is the Ia rate density\footnote{Note that we use a slightly higher value of the Ia rate (compared to previous studies), such that the net stellar heating is larger than the net radiative cooling. This value is still consistent with \cite{Maoz2017ApJ}. Furthermore, we test the effects of different Ia rates on our results in \S~\ref{subsec:diff_SNIa_rates}.}. The mass, momentum, and energy injection rate densities due to the SNIa are:
\begin{align}
    &\dot{\rho}_\mathrm{Ia} = 1~M_\sun\times \dot{n}_\mathrm{Ia},\label{eq:SNIa_massrate}\\
    &\dot{\bm{p}}_\mathrm{Ia} = 1~M_\sun \bm{v}_* \times \dot{n}_\mathrm{Ia},\label{eq:SNIa_momrate}\\
    &\dot{e}_\mathrm{Ia} = 10^{51}~\mathrm{ergs}\times \dot{n}_\mathrm{Ia},\label{eq:SNIa_heatrate}
\end{align}
respectively, assuming each SNIa event injects $1~M_\sun$ of mass and $10^{51}~\mathrm{ergs}$ of energy into the ISM, and $\bm{v}_*$ is the initial velocity of the progenitor system.

\subsection{Key timescales}\label{subsec:theory_time_scales}
\begin{figure}
		\centering
	\includegraphics[width=\columnwidth]{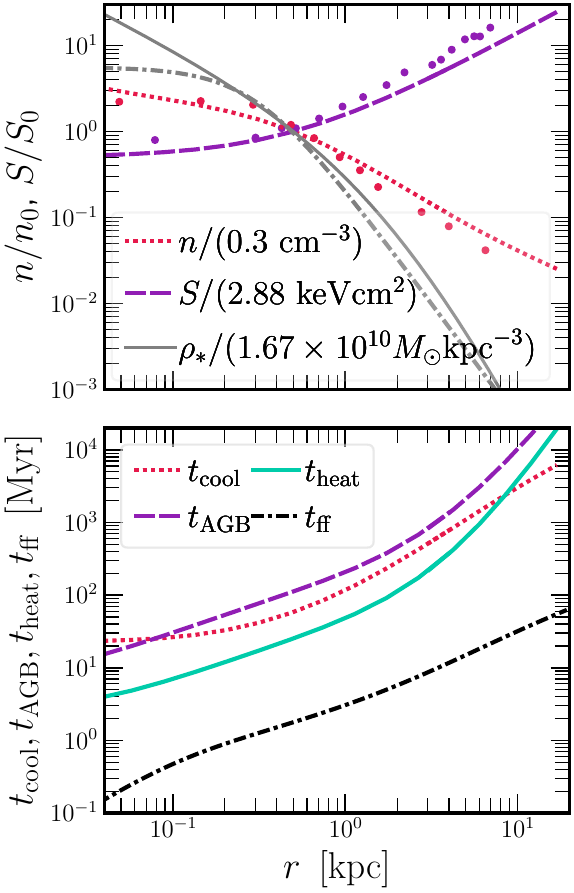}	
	\caption{Upper panel: The radial profiles of gas density (red dotted), entropy (purple dashed), and stellar density (grey solid). The scatter points show the profiles for the elliptical galaxy NGC 1399 from X-ray observations \citep{Werner2012MNRAS}. The grey dash-dotted line shows the stellar density distribution fit from \cite{Schuberth2010A&A}. \emph{Lower panel:} The radial profiles of important time-scales: the cooling time $t_\mathrm{cool}$, the AGB mass-injection time $t_{\mathrm{AGB}}$, the stellar heating time $t_\mathrm{heat}$, and the gravitational free-fall time $t_\mathrm{ff}$.}
	\label{fig:init_rad_prof_dens_entr_timescales}
\end{figure}

Associated with the source terms due to cooling and stellar evolution defined above are some key timescales relevant to our simulations. The radiative cooling time and the stellar heating time are given by:
\begin{subequations}
\begin{align}
    &t_\mathrm{cool} = e_\mathrm{int}/\mathcal{L}, \text{ and}\label{eq:t_cool}\\
    &t_\mathrm{heat} = e_\mathrm{int}/(\dot{e}_\mathrm{Ia}+\dot{e}_*), \label{eq:t_heat}
\end{align}
respectively, where $e_\mathrm{int}$ refers to the internal energy density of the gas.
The time it  takes for the AGB ejecta to replace the ISM is defined as:
\begin{equation}
    t_\mathrm{AGB} = \rho/\dot{\rho}_*, \label{eq:t_AGB}
\end{equation}
and finally the free-fall time is given by:
\begin{equation}
    t_\mathrm{ff} = \sqrt{2r/g(r)}, \label{eq:t_ff}
\end{equation}
where $g(r)$ is the acceleration due to gravity at radius $r$.
\end{subequations}

In the lower panel of Fig.~\ref{fig:init_rad_prof_dens_entr_timescales}, we show the radial profiles of the above timescales in our fiducial NGC 1399 initial condition. For $r\lesssim5~\mathrm{kpc}$, we find $t_\mathrm{ff}<t_\mathrm{heat}<t_\mathrm{cool}<t_\mathrm{AGB}$. As the stellar density falls off faster than the gas density (upper panel of Fig.~\ref{fig:init_rad_prof_dens_entr_timescales}), eventually $t_\mathrm{cool}\lesssim t_\mathrm{heat}$ for $r\gtrsim10~\mathrm{kpc}$. Therefore, we anticipate that the inner regions of the ISM will be impacted more by the effect of mass loss from AGB stars and the heating from type Ia supernovae. The inner ISM is expected to be completely replaced by the AGB ejecta in a few~$100~\mathrm{Myr}$, which is shorter than the cooling time at larger radii where Type Ia supernovae heating becomes less effective.   

%All of the other source terms have been already defined in Section~\ref{sec:theoretical_background}.
%\end{subequations}

\subsection{Numerical methods}\label{subsec:numerical_methods}
We evolve \crefrange{eq:continuity}{eq:energy} using \texttt{AthenaK}\footnote{\hyperlink{https://github.com/IAS-Astrophysics/athenak}{https://github.com/IAS-Astrophysics/athenak}}, a GPU-enabled, performance-portable version of the \texttt{Athena++} \citep{Stone2020ApJS,Stone2024arXiv}, implemented with the Kokkos library \citep{Trott2021CSE}. We employ second-order RK2 time integration, the HLLC Riemann solver and piece-wise linear spatial reconstruction. We also apply a first-order flux correction algorithm \citep{Lemaster2009ApJ} to address cells with unphysical velocities or temperatures.

\subsection{Heating and cooling implementation}\label{subsec:heating_cooling_implementation}
We implement AGB heating as a smooth source term, that follows the stellar density.   We model the supernovae as discrete energy (and mass) injection events. We divide the simulation domain into $24$ logarithmically-spaced radial bins. Then we evaluate the net stellar mass in each resulting spherical shell and for each time-step, we estimate the mean expected number of SNIa events using \cref{eq:SNIa_rate}. Following this, we generate the number of supernovae from a Poisson distribution with this mean and assign its location randomly within this shell. We also assign the remnant a random velocity $\bm{v}_*$ from a Gaussian distribution with width $\sigma_*$. We inject $10^{51}~\mathrm{ergs}$ of energy and $1~M_\sun$ mass uniformly within a spherical region of radius $r_\mathrm{Ia}$ ($\lesssim r_\mathrm{fade}$) around the location, where $r_\mathrm{fade}$ refers to the fade radius of the supernova bubble (where the supernova shock fades into a sound wave, as the pressure within the remnant matches the ISM pressure) and is given by:
\begin{equation}
    r_\mathrm{fade} = 17.3~\mathrm{pc} \, \left(\frac{n}{0.3~\mathrm{cm}^{-3}} \, \frac{T}{1.5\times10^7~\mathrm{K}}\right)^{-1/3}.\label{eq:R_fade}
\end{equation}
Resolving the fade radius is critical to capture the SNIa-driven shocks and their interaction with the ISM in the Sedov-Taylor phase of the remnant's evolution. We refer the reader to \cite{MLi2020ApJa} for a more detailed discussion on the validity of our injection method.

For a subset of our simulations, labeled `uni', we implement the heating due to the SNIa as a uniform heating term that follows the stellar density, similar to our AGB heating implementation.   This is not as physical but is more similar to what lower resolution simulations necessarily do and so is an instructive comparison point.

The radiative cooling rate per unit volume is given by:
\begin{equation}
    \mathcal{L} = n_H^2\Lambda(T),\label{eq:cooling}
\end{equation}
where $\Lambda(T)$ is a temperature-dependent cooling function from  \citep{Schure2009A&A} for $Z_\sun$ metallicity and $n_H=n\mu\chi_H$ denotes the hydrogen number density, where $\chi_H=0.715$ is the hydrogen mass fraction. We employ a floor temperature of $10^{4.2}~\mathrm{K}$.

\subsection{Boundary conditions}\label{subsec:boundary_conditions}
We fix the values of density and temperature to their initial values for radii beyond the outer boundary at $r=r_\mathrm{outer}$ (at 10 kpc for our fiducial model). The discussion in \S \ref{sec:theoretical_background} motivates why such a boundary condition is reasonable.
At the inner boundary, we implement a under-pressurized low-temperature sink (for $r\leq r_\mathrm{sink}$) with $T_\mathrm{sink}=10^{3.2}~\mathrm{K}$ and number density $n_\mathrm{sink}=5\times10^{-3}~\mathrm{cm}^{-3}$. Close to the sink, we linearly decrease $g$,  $\dot{\rho}_*$, $\mathcal{L}$, and $\dot{e}_*$ to $0$ for $r_\mathrm{sink}<r<2r_\mathrm{sink}$. We do not inject supernovae for $r<r_\mathrm{sink}$.

% For our runs with $r_\mathrm{sink}=5~\mathrm{pc}$, we start our simulations with $r_\mathrm{sink}=30~\mathrm{pc}$ and smoothly shrink it to $5~\mathrm{pc}$ over the first $10~\mathrm{Myr}$ of the simulation. We have verified that this shrinking procedure does not have any significant effects on the outcomes of our simulations.

\subsection{Mesh-refinement}\label{subsec:mesh_refinement}
We use static mesh refinement to achieve better spatial resolution in the inner regions, where the gas pressure is larger (and $r_\mathrm{fade}$ is smaller). We calculate the radial profile of $r_\mathrm{fade}$ at $t=0$ and ensure that both $r_\mathrm{fade}$ and $r_\mathrm{sink}$ are resolved by at least $4$ cells.

\subsection{List of simulations}\label{subsec:list_of_simulations}
\begin{table*}
    \centering
    \def\arraystretch{1.1}
    \caption{Simulation parameters and statistics for different runs}
    \label{tab:sim_params}
    \hspace{-1in}
    \resizebox{2.4\columnwidth}{!}{
        \begin{tabular}{lccccccccccr} 
            \hline
            Label                    & Base       & Refinement & $\Delta x_\mathrm{min}$ & SNIa     & SNIa rate                           & $E_\mathrm{Ia}$   & $n_0$                & $S_0$                           & $r_\mathrm{sink}$ & $r_\mathrm{outer}$ & $t_\mathrm{end}$\\
                                     & resolution & levels     & $(\mathrm{pc})$         & heating  & $(10^{10}M_\odot\mathrm{Myr})^{-1}$ & $(\mathrm{ergs})$ & $(\mathrm{cm}^{-3})$ & $(\mathrm{keV}~\mathrm{cm}^2)$  & $(\mathrm{pc})$   & $(\mathrm{kpc})$   &$(\mathrm{Myr})$ \\
            (1)                      & (2)        & (3)        & (4)                     & (5)      & (6)                                 & (7)               & (8)                  & (9)                            & (10)              & (11)               & (12)            \\            
            \hline                                                                                                                  
                                                                                                                                    
            % fid                      & $1024^3$   & $2$        & discrete & $440$                               & $10^{51}$         & $0.3$                & $2.88$                          & $5$               & $2.5$              & $200$           \\
            fid                      & $512^3$    & $4$        & $1.22$                  & discrete & $600$                               & $10^{51}$         & $0.3$                & $2.88$                          & $5$               &  $5$               & $200$           \\
            \hline                                                                                                                                                                                                                                      
            % $S_00.5$                 & $1024^3$   & $2$        & discrete & $440$                               & $10^{51}$         & $0.455$              & $1.44$                          & $5$               & $2.5$              & $110$           \\
            $S_00.75$-fix$P_0$       & $512^3$    & $4$        & $1.22$                  & discrete & $600$                               & $10^{51}$         & $0.395$              & $2.185$                         & $5$               & $5$                & $100$           \\
            $S_02.0$-fix$P_0$        & $512^3$    & $4$        & $1.22$                  & discrete & $600$                               & $10^{51}$         & $0.155$              & $5.80$                          & $5$               & $5$                & $200$           \\
            $S_04.0$-fix$P_0$        & $512^3$    & $4$        & $1.22$                  & discrete & $600$                               & $10^{51}$         & $0.086$              & $11.61$                         & $5$               & $5$                & $200$           \\
            $S_02.0$-fix$T_0$        & $512^3$    & $4$        & $1.22$                  & discrete & $600$                               & $10^{51}$         & $0.105$              & $5.80$                          & $5$               & $5$                & $100$           \\
            $S_04.0$-fix$T_0$        & $512^3$    & $4$        & $1.22$                  & discrete & $600$                               & $10^{51}$         & $0.037$              & $11.60$                         & $5$               & $5$                & $100$           \\
            \hline                                                                                                                                                                                                                          
            half-SN                  & $512^3$    & $4$        & $1.22$                  & discrete & $300$                               & $10^{51}$         & $0.3$                & $2.88$                          & $5$               & $5$                & $100$           \\
            double-SN                & $512^3$    & $4$        & $1.22$                  & discrete & $1200$                              & $10^{51}$         & $0.3$                & $2.88$                          & $5$               & $5$                & $100$           \\
            \hline                                                                                                                                                                                                                          
            bigSN10                  & $512^3$    & $4$        & $1.22$                  & discrete & $60$                                & $10^{52}$         & $0.3$                & $2.88$                          & $5$               & $5$                & $200$           \\
            bigSN64                  & $512^3$    & $4$        & $1.22$                  & discrete & $9.375$                             & $6.4\times10^{52}$& $0.3$                & $2.88$                          & $5$               & $5$                & $60$           \\
            \hline                                                                                                                                                                                                                      
            inner                    & $512^3$    & $3$        & $0.30$                 & discrete & $600$                               & $10^{51}$         & $0.3$                & $2.88$                          & $1$               & $0.625$            & $50$            \\
            \hline                                                                                                                                                                                                                                      
            % uni                      & $512^3$    & $3$        & uniform  & $440$                               & $10^{51}$         & $0.3$                & $2.88$                          & $5$               & $2.5$              & $200$           \\
            uni                      & $512^3$    & $4$        & $1.22$                  & uniform  & $600$                               & $10^{51}$         & $0.3$                & $2.88$                          & $5$               & $10$               & $200$           \\
            inner-uni                & $512^3$    & $3$        & $0.30$                 & uniform  & $600$                               & $10^{51}$         & $0.3$                & $2.88$                          & $1$               & $0.625$            & $50$            \\
            outer-uni$20$-long       & $512^3$    & $5$        & $2.44$                  & uniform  & $600$                               & $10^{51}$         & $0.3$                & $2.88$                          & $100$             & $20$               & $1200$          \\
            \hline                                                                                                                                                                                                                   
    \end{tabular}}
    \vspace{1.0em}
    \justifying \\ \begin{footnotesize} Notes: \emph{Column 1} indicates the simulation label. We show the base resolution, the number of refinement levels and $\Delta x_\mathrm{min}$ in \emph{Columns 2--4}. We denote the type of supernova heating implemented, the supernova event rate, and the energy injected per each event in \emph{Columns 5--7}. In \emph{columns 8--9} we list the number density ($n_0$), and the entropy ($S_0$) of gas at the entropy core radius $r_S$ at $t=0$. \emph{Columns 10 and 11} show the inner and outer boundary radii $r_\mathrm{sink}$ and $r_\mathrm{outer}$, in $\mathrm{pc}$ and $kpc$, respectively. Finally, we denote the duration of the simulation in $\mathrm{Myr}$ in \emph{column 12}. \end{footnotesize} 
\end{table*}
We have conducted 14 simulations to investigate various aspects of the impact of supernova Ia heating on massive galaxies, including:
\begin{enumerate}
    \item The effect of the initial entropy $S_0$ at $r=r_S$ (different $S_0$). Among these runs, there are two subsets, one where we keep the pressure at $r=2~\mathrm{kpc}$ fixed (`fix$P_0$' runs), and another where we keep the temperature at $r=2~\mathrm{kpc}$ fixed (`fix$T_0$' runs),% and the entropy slope $\alpha_S$ (different $\alpha_S$)
    \item The effect of different heating rates by varying the rate of injection of SNIa (`half-SN' and `double-SN') in our fiducial model,
    \item The impact of approximate modeling of supernovae, such as smoothly heating the ISM according to the expected SNIa rate (`uni'), clustered supernovae, where each supernova event has a larger energy, but the event rate is smaller to maintain a similar net heating rate (`bigSN10' and `bigSN64'),
    \item Following the accretion of matter down to the circularization radius close to the supermassive black hole (`inner') and (separately) following the long-term effects of Type Ia supernova and AGB mass-loss out to larger radii in the CGM, over $2$~$\mathrm{Gyr}$ (`outer-uni20-long').
\end{enumerate}
We list all of our simulations and some of their key parameters such as $n_0$, $S_0$, $r_\mathrm{sink}$, $r_\mathrm{outer}$ and $t_\mathrm{end}$ in \Cref{tab:sim_params}.

% ==================================
\section{Resolved Type Ia Supernova Feedback}\label{sec:resolved_sn1a_feedback}

\subsection{2D slices}\label{subsec:2d_proj_slice}
\begin{figure*}
		\centering
        \begin{interactive}{animation}{SN_radti_diff_S0_diff_T0.mp4}
        %figure call	
        \includegraphics[width=2\columnwidth]{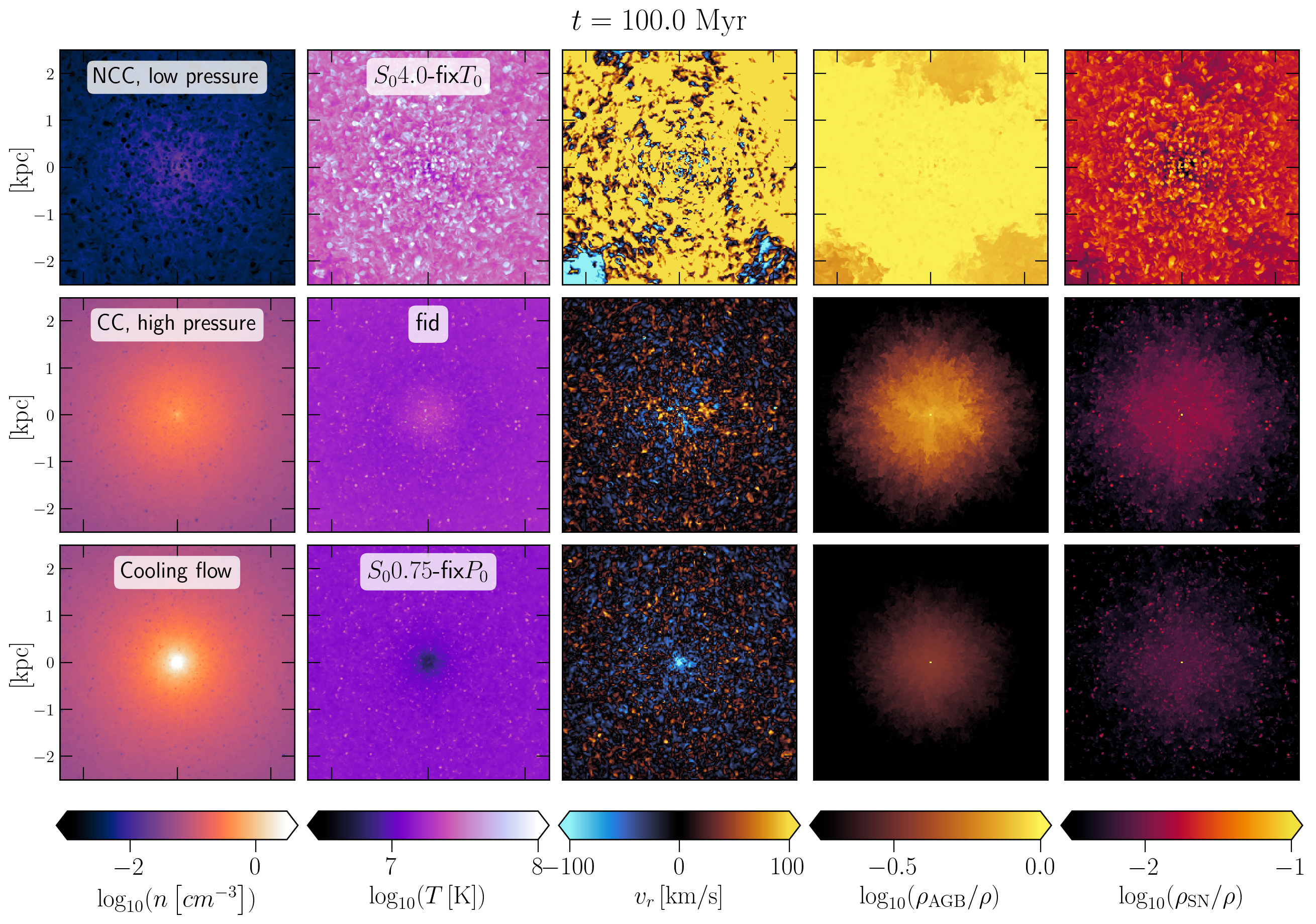}	
        \end{interactive}
	\caption{Slices (along the $xy$-plane) of number density (\emph{Col 1}), temperature (\emph{Col 2}), radial velocity (\emph{Col 3}), AGB ejecta mass fraction (\emph{Col 4}), and SNIa ejecta mass fraction (\emph{Col 5})  for the non-cool-core like `$S_04.0$-fix$T_0$' run, the cool-core like `fid' (\emph{Row 2}) run, and the low entropy `$S_00.75$-fix$P_0$' run (\emph{row 3}) at $t=100~\mathrm{Myr}$. The supernovae drive small-scale perturbations on $10$--$30~\mathrm{pc}$ scales. These three simulations represent the three classes of steady-state solutions shown in \Cref{fig:cartoon_diff_solutions}. An animated version of this figure, that captures the evolution from $0$ to $100~\mathrm{Myr}$ is available in the HTML version of the article.}% and on YouTube: \hyperlink{https://youtu.be/JvloSZhj0Mk}{https://youtu.be/JvloSZhj0Mk}.}}
	\label{fig:proj_slice_plot_fid}
\end{figure*}

In \Cref{fig:proj_slice_plot_fid}, we show  slices of number density (\emph{col 1}), temperature (\emph{col 2}), radial velocities (\emph{col 3}) and the mass fractions of stellar ejecta from AGB (\emph{col 4}) and SNIa (\emph{col 5}), respectively at $t=100~\mathrm{Myr}$ for the `$S_04.0$-fix$T_0$' (\emph{row 1}), `fid' (\emph{row 2}) and `$S_00.75$-fix$P_0$' (\emph{row 3}) runs. These three simulations represent the three classes of steady state solutions presented in \Cref{fig:cartoon_diff_solutions}:  `non cool-core', `cool-core', and cooling flow (from top to bottom, respectively).

From the density (\emph{col 1}) and temperature (\emph{col 2}) slices, we see that the top row is populated by several hot and under-dense bubbles/cavities inflated by the supernovae. The radial velocities (\emph{col 3}) show that Ia heating drives an outward wind, which takes most of the deposited mass and energy to larger radii. The inner $5~\mathrm{kpc}$ shown here is completely enriched by AGB ejecta, which reaches a mass fraction close to unity. The fraction of Ia ejecta reaches a few hundredths of the net mass and the SNe Ia ejecta are more volume filling than in the other simulations.

The `fid' run, shown in the middle row closely resembles the thermodynamic profiles of the NGC1399 galaxy from X-ray observations (\Cref{fig:init_rad_prof_dens_entr_timescales}). Since the surrounding ISM/ICM are at a larger pressure and lower entropy relative to the simulation in the top row, the Ia cavities are smaller in size ($r_\mathrm{fade}\propto P^{-1/3}$). The ISM is denser and cooler, and shows smaller variations compared to the lower pressure run. The radial velocities do not show a clear inflow/outflow structure, and are instead dominated by random velocities due to turbulence driven by the supernovae (there is in fact a net outflow outside $\sim 0.1$ kpc as we show in Figure \ref{fig:rad_prof_mflx_diff_S0} below but the mean velocity due to this outflow is smaller than the random velocity).  The enrichment of the ISM due to the AGB and SNIa ejecta is less over the 100 Myr simulation duration.
Although the AGB ejecta are injected smoothly with spherical symmetry, they show radial streaks similar to the SNIa ejecta, as they are transported to larger radii in the wakes of the buoyant Ia bubbles. 

Finally, the lowest entropy `$S_00.75$-fix$P_0$' run in the bottom row of Figure \ref{fig:proj_slice_plot_fid}
has a cooler and denser core compared to the `fid' run, with inflowing velocities in the central 1~$\mathrm{kpc}$ due to a cooling flow. Compared to the `fid' run, the fractional enrichment of the gas by AGB and SNIa ejecta is also smaller, likely due to the larger gas density and the net inflow of unenriched ISM from outer radii into the central $1~\mathrm{kpc}$. 

The three runs discussed here represent the three classes of steady-state solutions to the cluster density and entropy profiles that we showed in \Cref{fig:cartoon_diff_solutions}; we elaborate further on their properties below.

\subsection{Different initial thermodynamic profiles} \label{subsec:diff_S0}
\begin{figure*}
		\centering
	\includegraphics[width=1.7\columnwidth]{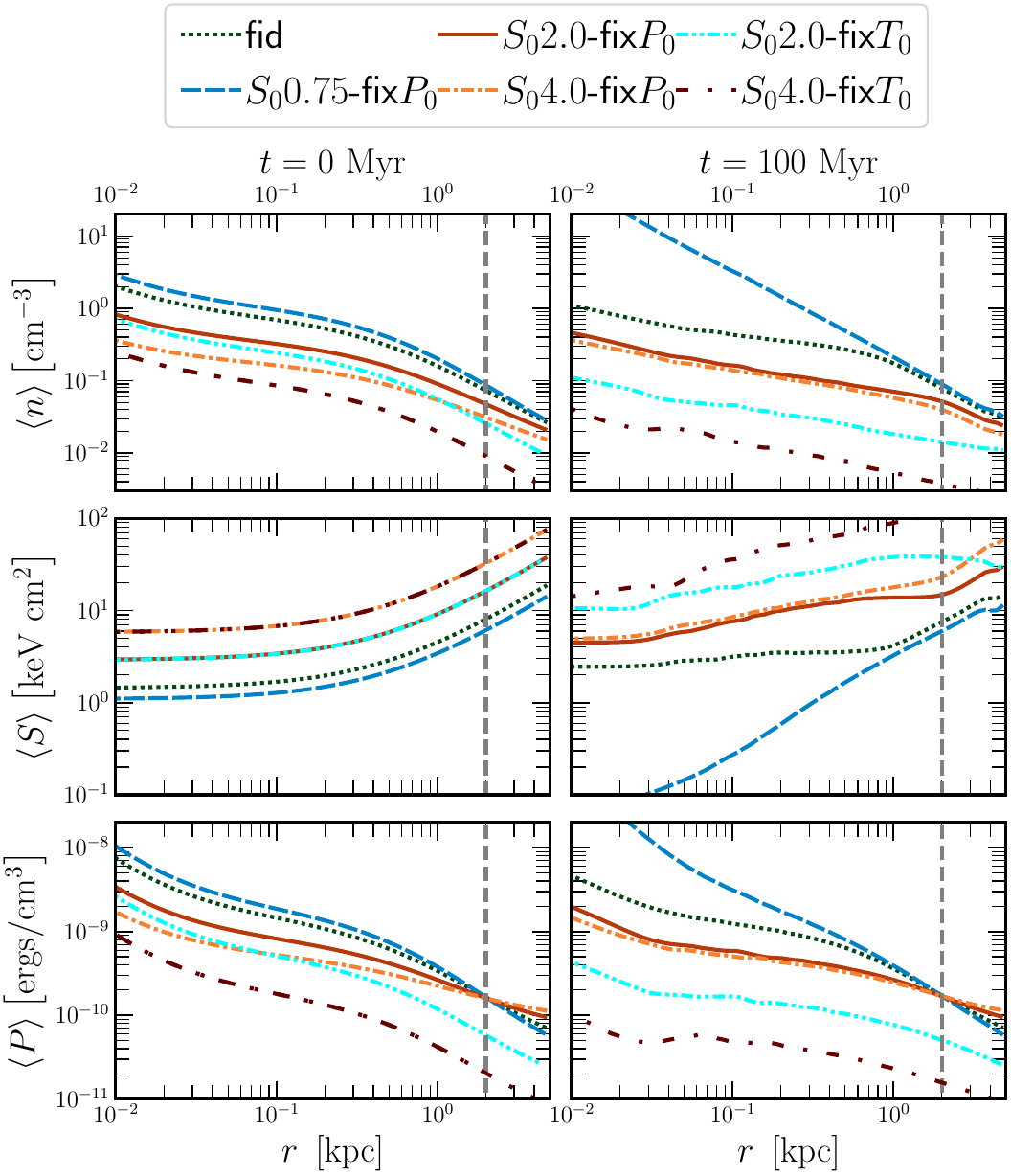}	
	\caption{Evolution of the radial profiles of gas density (first row), entropy (second row), and pressure (third row) for runs with different $S_0$ and either fixed pressure or temperature (at $r=2.0~\mathrm{kpc}$, denoted by the grey dashed line). For the solutions with similar pressure but different initial entropy, the density and entropy profiles evolve to a similar steady state profile for the `$S_02.0$-fix$P_0$' and `$S_04.0$-fix$P_0$' runs (relatively independent of initial conditions). The lower entropy `$S_00.75$-fix$P_0$' run, on the other hand approaches a different steady state solution (a cooling flow), with a much larger density and smaller entropy in the inner regions. The `fid' run shows a steady state profile in between these two solutions. The solutions with varying pressure (fixed $T_0$) differ as expected by the arguments in \S \ref{sec:theoretical_background} leading to the schematic Fig. \ref{fig:cartoon_diff_solutions}. }
	\label{fig:rad_prof_dens_entr_diff_S0}
\end{figure*}
\begin{figure}
		\centering
	\includegraphics[width=\columnwidth]{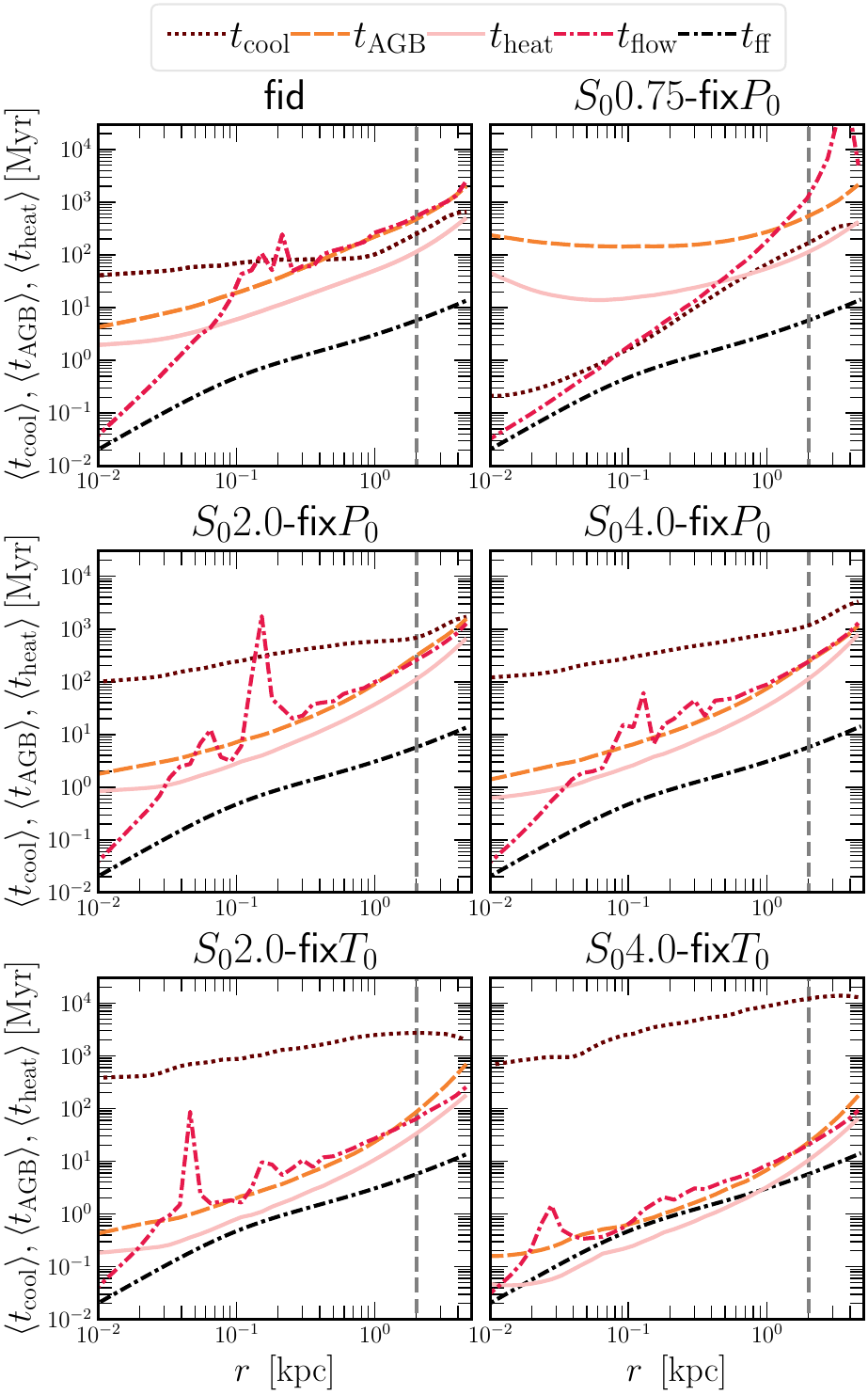}	
	\caption{Radial profiles of important timescales--$t_\mathrm{cool},t_\mathrm{AGB},t_\mathrm{heat}$, $t_\mathrm{flow} (=r/|v_r|)$, and $t_\mathrm{ff}$ for runs with different $S_0$, at $t=100~\mathrm{Myr}$. Except for the $S_00.75$ and the `fid' runs, the timescales are ordered as $t_\mathrm{ff}<t_\mathrm{heat}<t_\mathrm{AGB}<t_\mathrm{cool}$. For the `$S_00.75$' run, $t_\mathrm{cool}<t_\mathrm{heat}<t_\mathrm{AGB}$ for $r\lesssim0.5~\mathrm{kpc}$, which gives rise to a cooling flow in the inner regions, with $t_\mathrm{flow}\simeq t_\mathrm{cool}$ at $0.1~\mathrm{kpc}\lesssim r \lesssim0.5~\mathrm{kpc}$. For the `fid' run, $t_\mathrm{heat}<t_\mathrm{cool}<t_\mathrm{AGB}$. The peak in the radial profile of $t_\mathrm{flow}$ corresponds to the stagnation radius $r_\mathrm{stag}$, where the flow switches from an inflow at smaller radii to an outflow at larger radii.}
	\label{fig:timescales_prof_diff_S0}
\end{figure}
\begin{figure}
		\centering
	\includegraphics[width=\columnwidth]{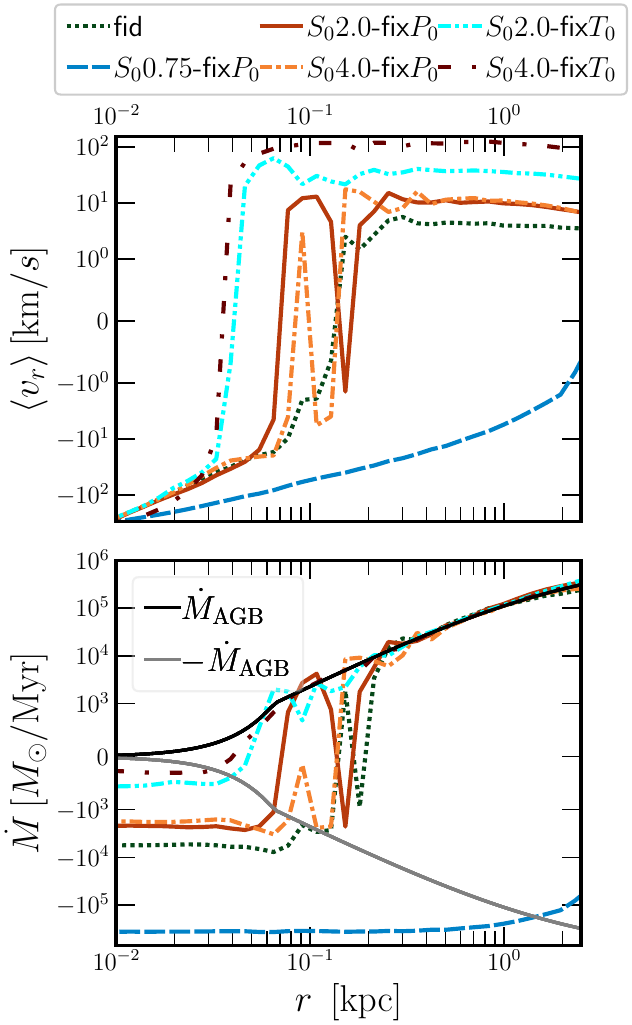}	
	\caption{The radial profiles of radial velocity (\emph{top row}) and mass flux (\emph{bottom row}): for runs with different $S_0$ at $t=100~\mathrm{Myr}$. The net mass injection rate due to AGB stars within a particular radius is denoted by $\dot{M}_\mathrm{AGB}$. The three higher entropy runs with similar ambient pressure show very similar velocity and mass flux profiles.   Decreasing the pressure (the `fix $T_0$' runs) leads to similar mass outflow rate but higher outflow speeds, consistent with \S \ref{sec:theoretical_background} and Fig. \ref{fig:cartoon_diff_solutions}.   The lowest entropy `$S_00.75$' run produces a strong cooling flow with larger inflow velocities and mass flux and a larger stagnation radius (where $\langle v_r\rangle=0$).  }
	\label{fig:rad_prof_mflx_diff_S0}
\end{figure}

In this subsection, we discuss the effects of our initial conditions on the radial profiles of density, entropy, and mass inflow/outflow rate. This parameter study helps us understand the different quasi steady-state profile solutions, and how they compare with the profiles inferred from X-ray observations. 
We conduct two sets of simulations, the first where we keep the pressure in the initial condition at $r=2$ kpc fixed and initialize four different entropy profiles with $S|_{r=r_S}=0.75S_0,\ S_0,\ 2S_0,\text{ and } 4S_0$. For the second set, we keep the initial temperature at $r=2$ kpc fixed, and initialize two different entropy profiles with $S|_{r=r_S}=2S_0,\text{ and } 4S_0.$   These are all different ways of exploring what range of solutions is produced by different initial gas properties and different ambient CGM/ICM pressures (at 2 kpc here%\footnote{Except for the `$S_02.0$-fix$P_0$' and `$S_04.0$-fix$P_0$' runs, where the outer boundary is at $2~\mathrm{kpc}$. However, the smaller outer boundary radius does not have any significant effects on the properties of the Ia driven wind.}
) for a given stellar distribution, Type Ia rate, and AGB mass-loss rate.     Of these simulations variations, those with fixed pressure at $r = 2~\mathrm{kpc}$ are the most similar to fixed ambient CGM pressure discussed in \S \ref{sec:theoretical_background}.

%\EQ{one thing i am confused by is exactly how this choice of profiles relates to our analytic arguments.   None of these choices is the same as keeping the pressure fixed at the outer boundary.   the ones with pressure at r = rS fixed is presumably closest to that, but it's not exactly that   so why are the solutions so strikingly similar even though the ambient pressure is actually changing by factors of a few  is it possible there is a bit of a coincidence and the solutions have only had time to reach AGB+Ia equilibrium out to a kpc not out to the outer boundary?}

%We set the outer boundary $r_\mathrm{out}$ at $5~\mathrm{kpc}$ to minimize any anomalous boundary interactions.\footnote{The outflows in the fiducial and the two higher entropy simulations reduce to a few $\mathrm{km/s}$ by $r_\mathrm{out}=2.5~\mathrm{kpc}$ and thus do not show strong interactions with the fixed boundary.}

We show the radial profiles of density, entropy, and pressure in \Cref{fig:rad_prof_dens_entr_diff_S0} at $t=0~\mathrm{Myr}$ (\emph{left col}) and $t=100~\mathrm{Myr}$ (\emph{right col}). Although the initial density and entropy profiles are different at $t=0~\mathrm{Myr}$, the three runs with fixed pressure `fid', `$S_02.0$-fix$P_0$' and `$S_04.0$-fix$P_0$' 
converge to a similar profile by $t=100~\mathrm{Myr}$. The fixed temperature runs `$S_02.0$-fix$T_0$' and `$S_04.0$-fix$T_0$' roughly maintain their initial density and entropy profiles.   The simulation that evolves the most in time is `$S_00.75$-fix$P_0$', in which the gas becomes much denser at small radii.

To better understand the simulation trends, we show the radial profiles of some of the important time-scales at $t=100~\mathrm{Myr}$ in \Cref{fig:timescales_prof_diff_S0}. Excluding the `$S_00.75$-fix$P_0$' run, and the `fid' run, the radial profiles of all other runs satisfy $t_\mathrm{ff}\lesssim t_\mathrm{heat}\lesssim t_\mathrm{AGB}\lesssim t_\mathrm{cool}$ (and so all represent versions of the left column class of solutions in Figure \ref{fig:cartoon_diff_solutions}).
Since $t_\mathrm{AGB}\lesssim100~\mathrm{Myr}$ for the `$S_02.0$-fix$P_0'$ and `$S_04.0$-fix$P_0$' runs, the gas in the inner $0.5$~$\mathrm{kpc}$ is replaced by the stellar ejecta within this time, with an effective entropy set by the stellar and supernovae energy deposited at the corresponding radii. However for the `$S_00.75$-fix$P_0$' run, $t_\mathrm{cool}\lesssim t_\mathrm{heat}\lesssim t_\mathrm{AGB}$ in the inner regions. As a result, it shows a cooling flow, with $t_\mathrm{flow}=r/v_r\simeq t_\mathrm{cool}$ around $r=0.1~\mathrm{kpc}$.    This is the origin of the much denser core and lower central entropy in Figure \ref{fig:rad_prof_dens_entr_diff_S0}. For the `fid' run, the timescales follow the order $t_\mathrm{heat}<t_\mathrm{cool}<t_\mathrm{AGB}$. The steady state thermodynamic profiles for this run lie in between the cooling flow and the `outflow' solutions, since the outflowing wind loses a large fraction of its energy to radiative cooling.

\Cref{fig:rad_prof_mflx_diff_S0} shows the radial velocity and mass flux profiles at $t=100~\mathrm{Myr}$ for all of the runs in Figures \ref{fig:rad_prof_dens_entr_diff_S0} \& \ref{fig:timescales_prof_diff_S0}. 
%Due to the lower confining pressure, the SNIa drive a stronger wind. 
%In comparison, the radial profiles of $\langle v_r\rangle$ are similar for the `fid' and fix$P_0$ runs. 
%For the fix$T_0$ runs, the core density is smaller for runs with higher entropy as the confining pressure decreases compared to the `fid' run. 
Except the cooling flow `$S_00.75$-fix$P_0$' run, all other runs show a similar flow structure--inflow for $r\lesssim$ some stagnation radius  $r_\mathrm{stag}$ and outflow at larger radii, where $r_\mathrm{stag}$ is defined as the radius where the flow switches from inflow-dominated to outflow-dominated. We find that $r_\mathrm{stag}$ in our simulations is smaller than the  $1~\mathrm{kpc}$ estimated in \citetalias{Voit2020ApJ} from their steady flow solutions. This difference could be due to their use of a lower SNIa rate in their calculations for NGC 1399\footnote{For effects of a different SNIa rate on our simulations, see \Cref{fig:rad_prof_dens_entr_mflx_diffSNrate}.}. %This is inline with expectations from the spherically symmetric steady state models of \cite{Generozov2015MNRAS}.   
The volume fraction of the ISM heated directly by the Ia remnants within a free-fall timescale is roughly given by $\dot{n}_\mathrm{Ia}V_\mathrm{Ia}t_\mathrm{ff}\approx 3\%\times (P/(4.5\times10^6 k_B\mathrm{cm}^{-3}\mathrm{K}))^{-1}$. Hence the stagnation radius is smaller for the runs with lower pressure (`fix$T_0$' runs) as they directly heat a larger fraction of the inner few $\mathrm{kpc}$. In addition, the lower bounding CGM pressure, roughly $3$ times smaller for the `$S_02.0$-fix$T_0$' run and $10$ times smaller for the `$S_04.0$-fix$T_0$' run, implies that Type Ia SNe can drive the AGB ejecta to larger velocities, as argued in eq. \ref{eq:vout} and indeed found in the simulations.  On the other hand, for the higher pressure `fix$P_0$' runs, the SNIa heat a smaller volume of the ISM directly, which rise buoyantly and deposit their energy further from the injection sites. The altered radial distribution of stellar energy injection results in a larger stagnation radius.  The net outflow rate is nearly the same in all models, however, since it is set by the  stellar mass return from AGB stars (bottom panel of Fig. \ref{fig:rad_prof_mflx_diff_S0}).   The different stagnation radii instead lead to a difference in how much of the AGB ejecta is accreted onto the massive black hole at small radii, as shown by the variation in the negative $\dot M$ in the bottom panel of Fig. \ref{fig:rad_prof_mflx_diff_S0}.

%We also show the mass deposition rate by the AGB stars $\dot{M}_\mathrm{AGB}$ within radius $r$ (and the negative of its value) for reference. In steady state, beyond $r_\mathrm{stag}$, the outflow mass flux matches $\dot{M}_\mathrm{AGB}$.

%The inflow $v_r$ is also similar for these three runs (for $r\lesssim r_\mathrm{stag}$). The inflow mass flux is set by the gas density and velocity at $r_\mathrm{stag}$ and has a similar amplitude for the `fid', `$S_02.0$-fix$P_0$' and `$S_04.0$-fix$P_0$' runs (which have nearly the same density at $r_\mathrm{stag}$). 

As with the thermodynamic profiles in Figure \ref{fig:rad_prof_dens_entr_diff_S0}, the simulation that is the most discrepant from the rest is `$S_00.75$-fix$P_0$', in which the gas undergoes a cooling-inflow with much larger velocities, since $v_r\approx r/t_\mathrm{cool}$ for a cooling flow. The inflow mass flux is roughly two orders of magnitude larger compared to the higher entropy runs.  %However, for $r\gtrsim1~\mathrm{kpc}$, $t_\mathrm{heat}<t_\mathrm{cool}$ and the gas becomes outflowing again, with both $v_r$ and $\dot{M}$ similar to the higher entropy runs. 

%The `fix$T_0$' runs have a smaller $r_\mathrm{stag}$, as well as smaller gas densities at $r_\mathrm{stag}$. Hence the gas inflow rates are much smaller for these runs compared to the other runs.

This section highlights that even when we initialize the galaxy with different density and entropy profiles, within a few~$~t_\mathrm{AGB}$, the entire inner region converges to a profile that depends primarily on the ambient CGM pressure and the properties of AGB mass loss and Type Ia supernova heating (so long as $t_\mathrm{AGB},t_\mathrm{heat}<t_\mathrm{cool}$, so that a cooling flow does not develop).   The system loses memory of other aspects of the initial condition of the simulation.   Simulations with low bounding pressure (higher entropy) lead to final profiles that are lower density and higher entropy -- this is possible for a fixed mass-return by stellar evolution because the outflow velocities are significantly larger (though still subsonic) in this case.  Thus Type Ia supernovae and AGB mass loss can maintain cool-core clusters as cool-core clusters and non-cool-core clusters as non-cool-core clusters.   Rather remarkably, the 
%, as long as $t_\mathrm{AGB},t_\mathrm{heat}<t_\mathrm{cool}$, and the confining CGM/ICM pressure is held fixed. The 
time-evolved profiles we find when we initialize the observed profiles of NGC 1399 (from X-ray data) remain  similar to the observations (our `fid' run). This implies that the observed profiles in the inner regions of massive galaxies are consistent with being set by the  (observed) confining CGM pressure and a combination of stellar mass loss and Type Ia SNe heating.   We discuss the role of AGN feedback in the context of these results in \S \ref{sec:Conclusion}.

\subsection{Implications for Heavy Elements and Dust}

The right two columns of Figure \ref{fig:proj_slice_plot_fid} show the distribution of the AGB and SNIa ejecta in the different classes of solutions found here.
% Cite observational papers regarding metallicity gradients here - for example 
Since the AGB ejecta are carbon, oxygen, and dust rich, and Ias are iron rich, the distributions in  Figure \ref{fig:proj_slice_plot_fid} are relevant for interpreting the gradients of metals, and the ratio between different elements, in X-ray observations,  eg.~\cite{Gatuzz2023MNRASa,Gatuzz2023MNRASb}.  They are also relevant for interpreting observations of PAH and infrared continuum emission from warm dust in several massive ellipticals \citep{Donahue2011ApJ}. Since dust molecules are expected to sputter within a few $\mathrm{Myr}$ in the hot ISM \citep{Draine2011piimbook}, they must be continuously injected into it through other mechanisms, such as from AGB ejecta (or pre-existing dust would need to be shielded from the hot ISM for many dynamical times).

For the fiducial run (middle row), we find the AGB ejecta mass fraction reaches almost $80$--$100\%$ for the inner $~\mathrm{kpc}$ and it decreases with radius for outer radii. This implies that the inner region is almost completely enriched by the stellar ejecta, and the weak outflow slowly adds metals to larger radii. This could explain why the inner regions of massive ellipticals exhibit$\approx$solar metallicity. We note here that we have assumed the AGB ejecta to completely thermalize with the ISM, which might not be always applicable to the high-pressure inner ISM of massive ellipticals \citep{YLi2019ApJ}, where the fast cooling mixing layers between the ejecta and the hot ISM can give rise to a long cool tail of AGB material that mixes inefficiently. 

In contrast to the AGB ejecta, the SNIa ejecta mass fraction peaks further out, close to $r\simeq 1~\mathrm{kpc}$ and reaches a peak of around $1\%$ since of course the mass-supply by AGB winds is much larger than that from SNIa. Since we inject the SNIa tracers only within the hot remnants, they rise buoyantly and mix with the ISM at larger radii from their injection sites.

We defer to future work a more detailed analysis of the implications of the passive scalar results in Figure \ref{fig:proj_slice_plot_fid} for observations of metals and dust in massive galaxies.   This is clearly a topic that merits a detailed study using simulations that resolve the ISM of massive galaxies, as we have done here.

%This result is similar to the metal-enriched outflows reported in galactic wind studies \citep{CGKim2020ApJ,Vijayan2024MNRAS}

%are also closest to the X-ray observed profiles of NGC 1399 which we used to initialize the `fid' run. So the thermodynamic profiles in the inner regions of the massive ellipticals are set by the confining pressure and a combination of stellar mass loss and heating. If other mechanisms such as AGN jets reduce the confining pressure at larger radii, the SNIa are able to drive a strong thermal wind, and maintain the thermodynamic profiles close to those of non-cool-core cluster in the inner few $\mathrm{kpcs}$.

\subsection{Effects of different Ia rates} \label{subsec:diff_SNIa_rates}
\begin{figure*}
		\centering
	\includegraphics[width=2.0\columnwidth]{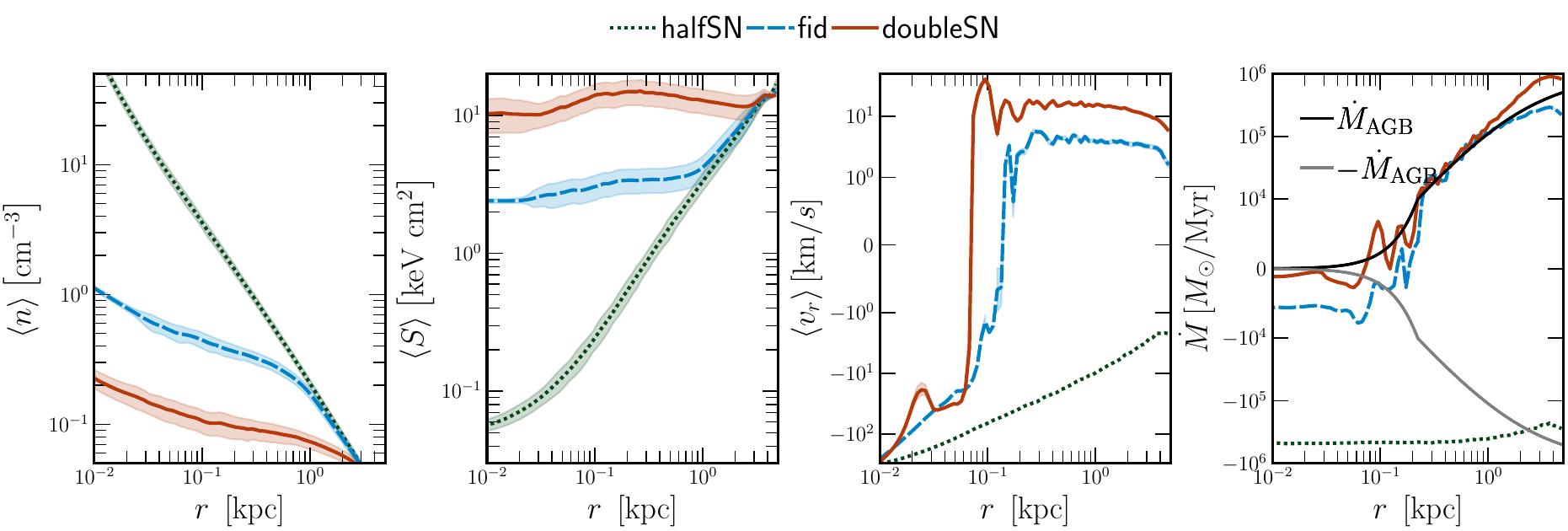}	
	\caption{The radial profiles of gas density (\emph{col 1}), entropy (\emph{col 2}), radial velocity (\emph{col 3}), and mass flux (\emph{col 4}) for different supernova injection rates `half', `fiducial' and `double', at $t=100~\mathrm{Myr}$. Stronger SNIa heating leads to a less dense, higher entropy core, drives a stronger outflow, and brings the stagnation radius closer to the black hole.  Weaker SNIa heating leads to a cooling flow.}
	\label{fig:rad_prof_dens_entr_mflx_diffSNrate}
\end{figure*}
Since there is some uncertainty in the estimated rate of type Ia supernovae, here we test the effects of different rates of SNIa. In \Cref{fig:rad_prof_dens_entr_mflx_diffSNrate}, we show the radial profiles of gas density  (\emph{col 1}), entropy (\emph{col 2}), radial velocity (\emph{col 3}), and mass flux (\emph{col 4}) for three different SNIa rates -- half of the fiducial rate, fiducial rate, and double the fiducial rate, at $t=100~\mathrm{Myr}$, all for our fiducial NGC 1399-motivated initial conditions. 

We find that increasing the SNIa rate decreases the gas density and increases the entropy in the central $1~\mathrm{kpc}$. The larger energy injection also drives a faster outflow. As expected, $r_\mathrm{stag}$ (where $\langle v_r\rangle=0$) is smaller for stronger heating, and the mass inflow rate onto the central black hole is smaller. The larger SNIa heating rate in the `double-SN' run overheats the core, and inverts the radial profile of entropy. We observe the onset of convection, and this simulation has not reached a statistical steady-state by $t_\mathrm{end}=100~\mathrm{Myr}$.  The simulation with half the fiducial Ia rate in \Cref{fig:rad_prof_dens_entr_mflx_diffSNrate} leads to a cooling flow in which a large fraction of the AGB mass loss is accreted by the central black hole.   
The results in \Cref{fig:rad_prof_dens_entr_mflx_diffSNrate} reinforce 
 our conclusion that the interplay between AGB mass loss and SNIa energy input establishes the density and entropy profiles on the scale of the stellar half-light radius.

\section{Approximate Type Ia Modeling} \label{sec:approx_1a_modeling}

\begin{figure*}
		\centering
	\includegraphics[width=2.0\columnwidth]{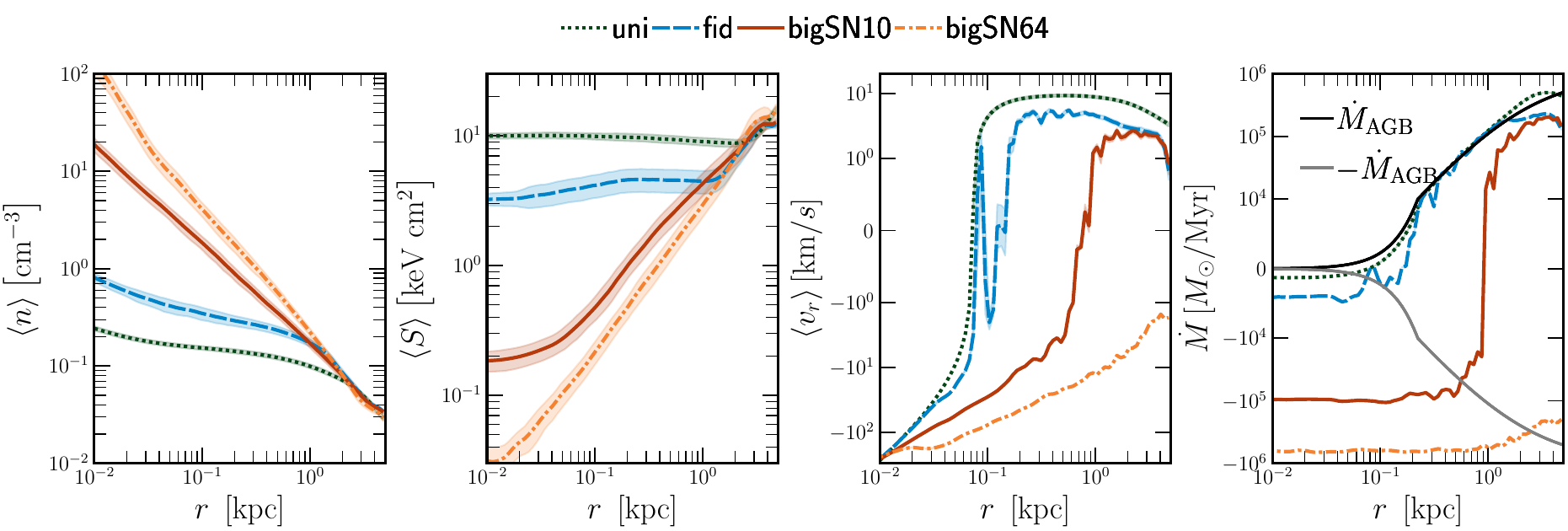}	
	\caption{The radial profiles of gas density (\emph{col 1}), entropy (\emph{col 2}), radial velocity (\emph{col 3}), and mass flux (\emph{col 4}) for different supernova injection methods `uni', `fiducial', `bigSN10', and `bigSN64', at $t=200~\mathrm{Myr}$. The larger bubbles generated my the clustered, bigger supernovae do not couple well to the inner ISM. So the ISM is cooler and denser and shows a cooling inflow for the `bigSN10' and `bigSN64' runs.}
	\label{fig:rad_prof_dens_entr_mflx_clusteredSNrate}
\end{figure*}

Modeling discrete SNIa in their Sedov-Taylor phase in the high-pressure central regions of massive ellipticals requires spatial resolution $\lesssim10~\mathrm{pc}$, since the individual remnants reach pressure equilibrium with their surroundings at  $\sim20~\mathrm{pc}$ (eq. \ref{eq:R_fade}). Large volume cosmological simulations often lack such high spatial resolution, and include their effects using sub-grid models \citep{Crain2015MNRAS,Pillepich2018MNRAS}.   Even Lagrangian zoom-in cosmological simulations would struggle to resolve the physics of Type Ia SNe remnants in hot gas given the low mass contained in a remnant at pressure equilibrium, $\sim 100 \, M_\odot \, (T/1.5\times10^7 \, {\rm K})^{-1}$ per eq.~\ref{eq:R_fade}.
In this section, we explore two approximate methods of numerical modeling of  remnants to compare to our `ground truth' resolved simulations.  

In the first approach (`uni' runs), we distribute the energy and mass injected by supernovae uniformly, following the stellar distribution. This method assumes that the energy injected by the supernovae is mixed into the ISM instantaneously. Many theoretical models such as \cite{Ciotti1991ApJ,Voit2015ApJ803L21V,Generozov2015MNRAS} and computational models such as \cite{CWang2019MNRAS}, that account for heating by SNIa also assume the heat is directly coupled to the interstellar medium in this manner, albeit with some efficiency parameter--these models are comparable to our `uni' simulations. These heating models neglect the energy redistribution caused by buoyantly rising bubbles, which as we show in this section, can lead to significant differences in their evolutionary outcomes.

In the second approach, we inject the SNIa as a cluster, such that each event is more energetic (by a factor of $10$ for `bigSN10', and $64$ for `bigSN64'), but the rate of such events is reduced to maintain the same heating rate.  This model is similar to the method implemented in the Eagle simulations \citep{Vecchia2012MNRAS,Schaye2015MNRAS}. Since the fade radius of the remnants $r_\mathrm{fade}\propto E_\mathrm{Ia}^{1/3}$, we need less spatial resolution to resolve the bigger supernovae. 

%\subsection{Radial profiles of density, entropy and mass fluxes}\label{subsec:rad_prof_dens_entr}

In \Cref{fig:rad_prof_dens_entr_mflx_clusteredSNrate}, we show the radial profiles of gas density  (\emph{col 1}), entropy (\emph{col 2}), radial velocity (\emph{col 3}), and mass flux (\emph{col 4}).

The `uni' run assumes that the hot gas within the SNIa remnants is instantaneously mixed with the ISM. Since $t_\mathrm{heat}\lesssim  t_\mathrm{cool}$ (see \Cref{fig:init_rad_prof_dens_entr_timescales}), the inner region of the ISM should approach a steady state temperature $\propto ((\dot{e}_\mathrm{Ia}+\dot{e}_*)/\dot{\rho}_*)^{1/2}$. Since $t_\mathrm{heat}$ is the shortest in the core, it is overheated compared to the outskirts. As a result, the low entropy gas in the core gets overheated, inverting the inner entropy profile. This gives rise to convective motions which flatten the entropy profile in the inner $\mathrm{kpc}$. 

For the `fid' run, the individual remnants are under-dense and rise buoyantly. They get mixed with the ISM during this process due to Rayleigh-Taylor and Kelvin-Helmholtz instabilities. As a result, they deposit their energy further out from their injection sites. Hence the core is no longer overheated, there is no $\mathrm{kpc}$ scale convection, and the entropy profile in the inner $1~\mathrm{kpc}$ does not flatten. The core is able to maintain its density and entropy structure for a much longer duration, close to the observed density and entropy profiles of nearby massive ellipticals.

We find that runs with clustered SNIa heating produce a denser and lower-entropy core compared to the fiducial run. Although the specific entropy of the remnants is similar to the fiducial run, the larger bubbles get disrupted over a longer time-scale ($\propto \sqrt{r_\mathrm{fade}/g}$). Thus, we expect the remnants to deposit their energy at larger radii compared to the smaller remnants in our `fid' run. Since the inner $\mathrm{kpc}$ core is no longer heated sufficiently by the bigger supernovae, it is denser and has a lower entropy, a larger $r_\mathrm{stag}$, and a larger black hole mass inflow rate. These effects are stronger for the `bigSN64' run, compared to the `bigSN10' run, and indeed the former leads to a cooling flow over most of the radii of interest.    The heating is too intermittent in the `bigSN64' case to effectively balance radiative losses.

% Some discussion on why this gas remains single phase and does not form any multiphase gas?
% Why is the temperature distribution so narrow?

% Here we discuss the effects of our two different heating models on radial velocities of the ISM and the mass accretion and ejection rates. 
We show the radial profile of the radial component of the gas velocity (\emph{col 3}) and the mass flux (\emph{col 4}) in Fig.~\ref{fig:rad_prof_dens_entr_mflx_clusteredSNrate}. We also show the mass deposition rate by the AGB stars within radius $r$ (and the negative of its value) for reference. The amplitude of inflow and outflow velocities is roughly similar for `uni' and `fid' runs. The gas is undergoing an inflow for $r\lesssim r_\mathrm{stag}\sim100~\mathrm{pc}$ and has a weak outflow with $v_\mathrm{out}\sim\text{ a few}~\mathrm{km/s}$ for $r>r_\mathrm{stag}$. The flow has a similar radial profile as the NGC 1399 steady-state solution of \citetalias{Voit2020ApJ}, albeit with a smaller $r_\mathrm{stag}$. 

For our discrete SNIa runs, we find that $r_\mathrm{stag}$ is somewhat larger compared to the `uni' run, since the supernova remnants rise buoyantly and deposit their energy at larger radii compared to the `uni' run.  The later stagnation radius also leads to a larger accretion rate onto the central black hole.
%As a result, only material further out from the center is able to escape from the gravitational pull of the black hole. This effect is particularly significant for the `bigSN64' run. The lack of core heating leads to a strong cooling flow in the inner $\mathrm{kpc}$. 

For all runs, the mass inflow rates are roughly independent of radius for $r\ll r_\mathrm{stag}$ down to the sink $\sim 10$ pc. The black hole accretion rate is few times larger than the Bondi accretion rate \citep{Bondi1952MNRAS}, and matches the net mass loss from AGB stars within $r_\mathrm{stag}$. 
%Hence, the inflow mass flux is larger for runs with larger $r_\mathrm{stag}$.
Further out, the outflow mass flux matches with $\dot{M}_\mathrm{AGB}$ for $r\gg r_\mathrm{stag}$--so the SNIa are able to sweep out the mass deposited by the AGB stars in the regions further away from the SMBH and keep the ISM in a steady state. The `bigSN64' run is  the exception, where the large remnants do not couple well to the ISM in the inner few $\mathrm{kpcs}$, leading to a cooling flow.

\section{Zoom-in and zoom-out simulations}\label{sec:zoom_in_zoom_out}
\begin{figure}
		\centering
	\includegraphics[width=1.\columnwidth]{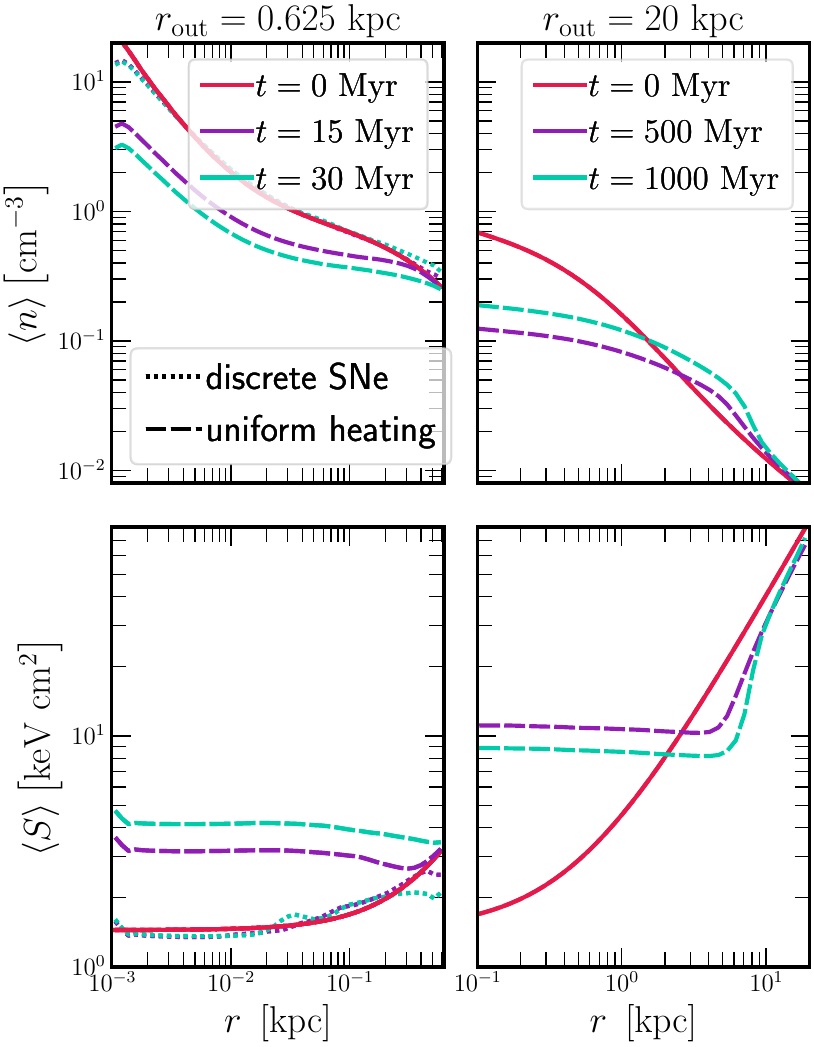}	
	\caption{Evolution of the radial profiles of gas density (first row) and entropy (second row) for runs with different $r_\mathrm{in}$ and $r_\mathrm{out}$. For the simulations with smaller sink radius (\emph{left column}), we observe similar trends in the evolution of the density and entropy profiles as the `fid' and `uni' set (\Cref{fig:rad_prof_dens_entr_mflx_clusteredSNrate}). For the `uni20' simulation (\emph{right column}) the density and entropy profiles start showing evidence of a cooling inflow at larger radii due to insufficient stellar heating at $r\gtrsim10~\mathrm{kpc}$.}
	\label{fig:rad_prof_dens_entr_diff_r_out}
\end{figure}

%\EQ{maybe something about random J here?}

In this section, we discuss the dependence of our results on our choice of sink radius and outer boundary.  The simulation with a larger outer boundary is also run for a much longer timescale of $2~\mathrm{Gyr}$. The gas density and entropy profiles for these runs are shown in \Cref{fig:rad_prof_dens_entr_diff_r_out}. 

In the left column of \Cref{fig:rad_prof_dens_entr_diff_r_out} we show the results for simulations with $r_\mathrm{in}=1~\mathrm{pc}$ and $r_\mathrm{out}=0.625~\mathrm{kpc}$ at $t=15$ and $30~\mathrm{Myr}$, for both discrete SNe and uniform heating.   The density and entropy profiles remain almost unchanged for the discrete SNe simulation, whereas for SNe modeled via uniform heating, the density decreases and the entropy increases by a factor of $2$--$3$ within the first $30~\mathrm{Myr}$. These trends are similar to our main results discussed in \S \ref{sec:resolved_sn1a_feedback} and \S \ref{sec:approx_1a_modeling}, and show that the choice of inner sink radius does not have a strong effect on our results.   The reason is that so long as the sink radius is well inside the stagnation radius the dynamics are not sensitive to the exact value of the sink radius.    It is worth noting that the profiles at small radii in the left column of \Cref{fig:rad_prof_dens_entr_diff_r_out} are essentially those of spherical Bondi accretion and the constant entropy is because the inflow time (of order the free-fall time) is short compared to the cooling time and Ia heating timescale. We have checked that the random angular momentum generated by the SNe Ia-generated turbulence is relatively small and does not lead to significant angular momentum support on the scales simulated here.

We show the long-term evolution of the gas profiles in the right column of \Cref{fig:rad_prof_dens_entr_diff_r_out} for the simulation with $r_\mathrm{in}=100~\mathrm{pc}$ and $r_\mathrm{out}=20~\mathrm{kpc}$. We use uniform Ia heating here and do not perform a discrete SNIa-heating counterpart of this run, since it is computationally challenging to resolve $10$--$50~\mathrm{pc}$ size bubbles over a $(40~\mathrm{kpc})^3$ volume. We also expect the discreteness of SNIa to affect the inner few $\mathrm{kpc}$ regions more, where the stellar density is the highest. \Cref{fig:rad_prof_dens_entr_diff_r_out} shows that the gas density decreases and the entropy profile flattens to a higher value in the first $500~\mathrm{Myr}$. These results are similar to what we reported in \S\ref{sec:approx_1a_modeling} for the `uni' run, and would likely be somewhat different for a simulation with discrete SNe (as the comparison of discrete vs. uniform heating in Figure \ref{fig:rad_prof_dens_entr_mflx_clusteredSNrate} shows).   More interesting is that in the second $500~\mathrm{Myr}$ in the larger volume longer duration simulation, cooling of gas at large radii ($r\gtrsim10~\mathrm{kpc}$) leads to the early stages of a cooling flow, which increases the gas density and lowers the gas entropy compared to the values at $t=500~\mathrm{Myr}$ (the radial velocity is also negative at large radii, i.e. the SNIa-driven outflow has become a cooling-driven inflow). Note from \Cref{fig:init_rad_prof_dens_entr_timescales} that in the initial condition, $t_\mathrm{cool}\sim t_{\rm heat} \sim 3 ~\mathrm{Gyr}$ at $r\sim 10 ~\mathrm{kpc}$; it is thus not obvious that Ias alone can suppress cooling at large radii and indeed they do not appear to do so in our longer duration simulation. %Even in the absence of cooling, the specific energy of the outflow is insufficient to drive the gas beyond $r\sim100~\mathrm{kpc}$, also argued in \citetalias{Voit2020ApJ}. 
This  highlights the potential importance of AGN in heating the gas at larger radii, where stellar heating is insufficient to offset the cooling losses; we return to this point in the next section.
\section{Summary and Discussion}\label{sec:Conclusion}
In this paper we used hydrodynamic simulations to study the central few $\mathrm{kpc}$ of massive elliptical galaxies. We  focused on the effect of  heating due to type Ia supernovae (SNIa) and mass return to the ISM from evolved stars, primarily those on the asymptotic giant branch (AGB).   The motivation for doing so is that theory and observations show that the timescale for SNe to heat the ISM and the timescale for AGB mass loss to replenish the ISM are both quite short, $\sim 100$ Myr (Fig. \ref{fig:cartoon_diff_solutions}).   SNe Ia and AGB mass loss must thus play an important role in the evolution of the ISM of massive galaxies (e.g., \citealt{Mathews1971ApJ,Mathews1986ApJ,Ciotti1991ApJ,Negri2014MNRAS445,Conroy2015ApJ,Generozov2015MNRAS,Voit2015ApJ803L21V,Voit2020ApJ}).   
%We are motivated by the fact that the SNe
%is comparable to the net radiative cooling of the ISM in these regions of many nearby ellipticals, we have studied their effects on the dynamics and evolution of the ISM. We have tested the effects of different initial entropy profiles, different SNIa rates, and compared between modeling the SNIa as discrete events versus approximate methods used in cosmological simulations, such as clustered supernovae and a spatially smooth heating function. We have also looked into the effect of our choice of different inner and outer radii on the evolution of the ISM.
Our key findings are summarized below, after which we discuss some of the implications and limitations of our study.

%\begin{enumerate}

We consider simulations with a range of different initial density and temperature  profiles, with our fiducial model motivated by observations of NGC 1399.  We also consider variations in the Type Ia supernova rate per unit stellar mass of a factor of $\sim 4$.   In doing so, we find that two different quasi-steady state solutions emerge (\Cref{fig:cartoon_diff_solutions} and \Crefrange{fig:proj_slice_plot_fid}{fig:rad_prof_mflx_diff_S0}):

\

\noindent 1.  When the heating time of the ISM due to Type Ia supernovae $t_{\rm heat}$ is shorter than the cooling time $t_{\rm cool}$,  the type Ia supernovae drive a hot wind out into the CGM. This outflow drives away almost all the material injected by AGB winds, with only a small fraction of the AGB ejecta accreted by the central black hole. 
        %The core resembles a non-cool-core, with entropy $\gtrsim10~\mathrm{keVcm}^2$.
A key property of this solution is that the SNIa-driven wind is technically not a supersonic wind, but a subsonic `breeze'.   Because the outflow is subsonic  some of its properties depend on the confining pressure of the CGM exterior to the galaxy at a few effective radii $\sim {\rm few}-10$ kpc.  When the confining pressure due to the CGM/ICM is low, the Ia-driven wind has a higher speed and the ISM on galaxy scales is `non-cool-core' like, at a lower pressure, lower density and higher entropy.   By contrast, when the confining pressure of the CGM/ICM is higher, the outflow velocity is significantly lower and the ISM on galaxy scales is `cool-core' like, at a higher density and lower entropy.   Effectively, the pressure confinement of the Ia-driven AGB wind ensures that the ISM on galaxy scales inherits the rough thermodynamic properties of the surrounding CGM (e.g., high or low entropy) even when nearly the entire ISM on galaxy scales has been replaced by AGB ejecta, in agreement with the predictions of \citetalias{Voit2020ApJ}.   We note that this  result is specific to the low ram pressure of Ia-driven winds and the high ambient CGM pressure  in massive galaxies (\S \ref{sec:theoretical_background}).   In rapidly star-forming galaxies, by contrast, the galactic wind is almost always initially highly overpressured compared to the inner CGM and so drives a shock into the CGM (i.e, the wind modifies the CGM, not the CGM modifies the wind!). 

% Our 3D hydrodynamic simulations of resolved type Ia supernovae+AGB injection are in agreement with both the low and high confining CGM pressure solutions presented in (\citetalias{Voit2020ApJ}) for stellar heating in a galaxy with a deep potential well ($\sigma_*\gtrsim300~\mathrm{km/s}$). 

\               

\noindent 2.   If $t_\mathrm{cool}<t_\mathrm{heat},t_\mathrm{AGB}$, a second solution emerges, where the ISM is inflowing. The core is denser and cooler, and resembles a cooling flow with a low core entropy $\simeq0.1~\mathrm{keVcm}^2$. 

\

%The first class of solutions described above is a quasi-steady state solution in the sense that the solution does not evolve significantly over 100s Myrs, which is many dynamical times in the inner few kpc of massive galaxies.   However, in most cases the physics included here is likel

When we vary the rate of Type Ia supernovae per unit stellar mass for our fiducial NGC 1399 model (while keeping the initial thermodynamic profiles and the bounding pressure fixed), we find that the Ia rate we have used (consistent within $\sim 1 \sigma$ of the results of \citealt{Maoz2017ApJ}) leads to density and temperature profiles in good agreement with those observed in NGC 1399 (Figs. \ref{fig:init_rad_prof_dens_entr_timescales} \&  \ref{fig:rad_prof_dens_entr_diff_S0}).    Increasing the type Ia rate by a factor of 2 increases the core entropy, decreases the core gas density and drives a higher speed (but still subsonic) outflow.  By contrast, reducing the Type Ia rate by factor of 2 gives rise to a denser, cooler core with a cooling flow in the center (\Cref{fig:rad_prof_dens_entr_mflx_diffSNrate}).  NGC 1399 is clearly on the boundary between the two classes of solutions enumerated above, consistent with its location in the X-ray luminosity and Type Ia heating rate vs stellar mass plot in Figure \ref{fig:MX_Mstar_Q_SNIa}.

Although we intentionally do not include BH feedback in our simulations, we do include the gravity of a central $10^9 M_\odot$ BH that dominates the gravity in the inner $\sim 100$ pc, a region well-resolved in most of our simulations.   Not surprisingly the accretion rate onto the black hole varies strongly across the range of ISM solutions enumerated above.   The black hole accretion rate is an increasing function of the bounding CGM pressure.  It is largest for the cooling flow in which most of the AGB ejecta from the galaxy accrete onto the black hole, and is nearly a factor of $\sim 30$ smaller for even our low entropy cool-core like model which has $\dot M \sim 0.005 M_\odot {\rm yr^{-1}}$.    The black hole accretion rate is another factor of $\sim 10$ smaller for our highest entropy non-cool-core like model.   It is likely that the accretion rates estimated here are upper limits on the true accretion rate.   The reason is that although our simulations are three dimensional, the accretion flow onto the black hole is nearly spherical, with only a small angular momentum from the turbulent velocity generated by intermittent Type Ia supernovae.  It is likely that there is larger angular momentum on galaxy scales from other processes (black hole feedback itself, structure formation, satellite galaxies, small but non-zero rotation of the stars, etc).   In radiatively inefficient accretion flows with significant angular momentum, the accretion rate onto the black hole is substantially reduced relative to the Bondi rate found here \citep{Cho2023,MGuo2023ApJ}.

Because the focus of this study is simulations that resolve individual Ia remnants, we are unable to run most of our simulations long enough to understand the many Gyrs evolution of the system. What we instead find are quasi-steady states that are produced and do not evolve significantly over several hundred Myrs.   In one case, a simulation with a larger sink radius and uniform Type Ia heating (to decrease the resolution requirements) we ran our simulation for  
$1.2~\mathrm{Gyr}$ (comparable to the cooling time at $r\gtrsim10~\mathrm{kpc}$).  For the first 0.5 Gyr the solution is very similar to that described above with a slow outflow of AGB ejecta.   However towards the end of the simulation, the system starts to undergo a cooling flow once the gas beyond $10~\mathrm{kpc}$ cools.   Given sufficient time this would develop into a cooling flow at smaller radii as well.   This result is unsurprising since it is well known that stellar feedback cannot balance radiative losses in the most massive halos, particularly at larger radii where cooling peaks in the CGM (Fig. \ref{fig:MX_Mstar_Q_SNIa}).   In many isolated elliptical galaxies this is unlikely to be an issue and the Type Ia-regulated ISM found here may describe the system for the bulk of its lifetime.   For the most massive halos, however, another energy source is required, presumably BH feedback. Our results are in agreement with \cite{Ciotti1997ApJ,Ciotti2001ApJ,CWang2019MNRAS}, who showed that intermittent AGN activity is required to keep the ISM hot, even for single-phase elliptical galaxies, similar to the one we model here.

Our results provide some insight into how Type Ia supernovae, AGB ejecta, and BH feedback may work together to produce stable long-lived systems without cooling flows; they are consistent with and support the basic picture proposed in \citet{Voit2020ApJ}.   In particular, a natural worry raised by the hypothesis that Type Ia supernovae are important on galaxy scales even in many massive groups and clusters is that this will break the necessary feedback loop that couples BH feedback and  accretion to the surrounding CGM properties.  This is not in fact the case.  Because the galaxy-scale ISM is pressure confined by the surrounding CGM, the galaxy scale properties including the BH accretion rate are directly influenced by the CGM pressure at large radii where the cooling flow problem is the most severe.   It is thus possible (though certaintly not necessary) that BH feedback's primary role is to heat the CGM at large radii $\sim 10s$ kpc where the cooling inflow rate is the largest (e.g., physically this could be because jets deposit their energy in the CGM not on kpc-scales).  We have shown that this is still fully compatible with a stable feedback loop between the CGM and BH, communicated via the pressure-confined ISM on galaxy scales.  Indeed, Type Ia supernovae may help maintain the stability of this feedback loop by suppressing cooling on $\sim 100$ Myr timescales at radii $\sim$ kpc, as found here.    A clear next step in the research presented here is carry out simulations with Type Ia SNe, AGB mass loss, and BH feedback to better understand the interplay between these three processes.

%he inner ISM ($r\lesssim0.1~\mathrm{kpc}$) is mostly inflowing, whereas the outer ISM has a weak outflow velocity $\lesssim10~\mathrm{km/s}$. Our runs with different density and entropy profiles converge to the same profile within $100~\mathrm{Myr}$. The converged values are close to the X-ray observed density and entropy profiles and are set by the combination of mass and energy injection due to AGB stars and type Ia supernovae. We denote these as cool-core-like solutions.
    
All of our primary simulations were tailored to explicitly resolve Type Ia SN remnants, which have sizes of $\sim 20$ pc and swept up masses of $\sim 100 M_\odot$ when they reach pressure equilibrium with the hot CGM.    Most global simulations do not have the resolution to do so.   As part of our study, we also explored two different numerical methods to approximate the effect of SNIa heating when individual remnants can't be resolved -- (i) uniform deposition of heat due to the SNIa following the stellar distribution and (ii) clustered SNIa injection where individual SNIa events are more energetic (and hence larger in pressure equilibrium), but less frequent. These approximate methods show  several key differences compared to our fiducial models with individually resolved remnants.

For uniform heating, the inner ISM is overheated compared to the fiducial simulations with resolved remnants; the core is higher entropy and low density, at roughly $5$-times the observed value in NGC 1399 (see \Cref{fig:rad_prof_dens_entr_mflx_clusteredSNrate}). In the case of individually resolved remnants, much of the Type Ia energy deposition is non-local as buoyant bubbles rise and mix, depositing their energy at larger radii than where the SNe originate.   This cannot be modeled in simulations with uniform heating that is proportional to the local stellar density.   Models with `clustered' Type Ia SNe suffer the opposite problem.   The buoyant bubbles are larger and do not get disrupted as easily as individual Type Ia remnants do. These larger bubbles do not heat the ISM efficiently in the central few $\mathrm{kpc}$). Indeed too much clustering of the SNe leads to a cooling flow.    Because the ISM properties on galaxy scales depend on the details of Type Ia heating, approximate methods of treating SNIa also lead to incorrect predictions for the black hole accretion rate: the uniform heating simulations underpredict the black hole accretion rate while the clustered SNe simulations overpredict it.

There are many other important extensions to the work presented here that would improve the fidelity of our models of massive galaxies.   Although we keep track of AGB and SNIa ejecta in our present simulations, we do not include their effect on the cooling rate of the ISM. The ejecta-rich regions are expected to undergo faster cooling, and these regions could form cold gas from condensation of the hot ISM.  We have also assumed the AGB material to be completely thermalized with the ISM, although \citet{YLi2019ApJ} show that the ejecta material might survive in the dense, high-pressure central regions of massive ellipticals. We plan to include a cooling function that depends on the ejecta self-consistently in future studies.   It would also be valuable to study a range of stellar density profiles bracketing observed systems (e.g., cores and cusps, models with intracluster light, etc).  Since the SNIa in the central regions rise up buoyantly and deposit their energy further out from where they originate, we do not expect a different stellar density profile to have a significant effect on the main results of this work.   It would nonetheless be useful to explicitly study this in more detail.

\section{Acknowledgements} 
% \begin{acknowledgments}

% The authors would like to thank the anonymous referee for a constructive report, which helped to improve this work.
% RM would like to thank Minghao Guo for help with setting up the simulations using Athenak and post-processing the results. 
RM thanks Joop Schaye and Greg Bryan for useful discussions.   We thank Mark Voit and Yuan Li for useful comments on the paper and the referee for very helpful comments and questions.  This work was supported in part by a Simons Investigator award from the Simons Foundation (EQ) and by NSF grant AST-2107872. It was supported in part by grant NSF PHY-2309135 to the Kavli Institute for Theoretical Physics (KITP) and NSF grant PHY-2210452 to the Aspen Center for Physics.
The analysis presented in this article was performed in part on computational resources managed and supported by Princeton Research Computing, a consortium of groups including the Princeton Institute for Computational Science and Engineering (PICSciE) and the Office of Information Technology’s High Performance Computing Center and Visualization Laboratory at Princeton University.
We also used the Delta GPU machine at National Center for Supercomputing Applications, Illinois, United States through allocations PHY230106 and PHY230045 from the Advanced Cyberinfrastructure Coordination Ecosystem: Services and Support (ACCESS) program, which is supported by National Science Foundation grants 2138259, 2138286, 2138307, 2137603, and 2138296. The Delta research computing project supported by the National Science Foundation (award OCI 2005572), and the State of Illinois. Delta is a joint effort of the University of Illinois at Urbana-Champaign and its National Center for Supercomputing Applications.
We acknowledge the EuroHPC Joint Undertaking for awarding this project access to the EuroHPC supercomputer LUMI, hosted by CSC (Finland) and the LUMI consortium through a EuroHPC Regular Access call.
% \end{acknowledgments}

% This work used the following software/packages:
\software{\texttt{AthenaK} \citep{Stone2020ApJS,Stone2024arXiv}, \texttt{matplotlib} \citep{Hunter4160265}, \texttt{cmasher} \citep{Ellert2020JOSS}, \texttt{scipy} \citep{Virtanen2020}, \texttt{NumPy} \citep{Harris2020}, \texttt{CuPy} \citep{Okuta2017CuPyA}, \texttt{h5py} \citep{collette_python_hdf5_2014}, \texttt{LLaMA} \citep{LLamaMetaAI}, and \texttt{astropy} \citep{astropy2018}}.

%%%%%%%%%%%%%%%%%%%%%%%%%%%%%%%%%%%%%%%%%%%%%%%%%%

\section{Data Availability}
All relevant data associated with this article is available upon reasonable request to the corresponding author.

\section{Additional Links}
Movies of our simulations are available at the following links on Youtube:
\begin{itemize}
    \item Effects of different initial thermodynamic profiles: \hyperlink{https://youtu.be/59lharS2CAc}{https://youtu.be/59lharS2CAc},
    \item Effects of different type Ia supernova rates: \hyperlink{https://youtu.be/1NLRuS5xT-U}{https://youtu.be/1NLRuS5xT-U},
    \item Effects of different methods of implementing type Ia supernovae heating: \hyperlink{https://youtu.be/aF9vinVhDv0}{https://youtu.be/aF9vinVhDv0}.
\end{itemize}
%%%%%%%%%%%%%%%%%%%% REFERENCES %%%%%%%%%%%%%%%%%%

% The best way to enter references is to use BibTeX:

\bibliographystyle{aasjournal}
\bibliography{refs.bib} % if your bibtex file is called example.bib

\begin{thebibliography}{}
\expandafter\ifx\csname natexlab\endcsname\relax\def\natexlab#1{#1}\fi
\providecommand{\url}[1]{\href{#1}{#1}}
\providecommand{\dodoi}[1]{doi:~\href{http://doi.org/#1}{\nolinkurl{#1}}}
\providecommand{\doeprint}[1]{\href{http://ascl.net/#1}{\nolinkurl{http://ascl.net/#1}}}
\providecommand{\doarXiv}[1]{\href{https://arxiv.org/abs/#1}{\nolinkurl{https://arxiv.org/abs/#1}}}

\bibitem[{{Anderson} {et~al.}(2015){Anderson}, {Gaspari}, {White}, {Wang}, \& {Dai}}]{Anderson2015MNRAS}
{Anderson}, M.~E., {Gaspari}, M., {White}, S. D.~M., {Wang}, W., \& {Dai}, X. 2015, \mnras, 449, 3806, \dodoi{10.1093/mnras/stv437}

\bibitem[{{Astropy Collaboration} {et~al.}(2018){Astropy Collaboration}, {Price-Whelan}, {Sip{\H{o}}cz}, {G{\"u}nther}, {Lim}, {Crawford}, {Conseil}, {Shupe}, {Craig}, {Dencheva}, {Ginsburg}, {Vand erPlas}, {Bradley}, {P{\'e}rez-Su{\'a}rez}, {de Val-Borro}, {Aldcroft}, {Cruz}, {Robitaille}, {Tollerud}, {Ardelean}, {Babej}, {Bach}, {Bachetti}, {Bakanov}, {Bamford}, {Barentsen}, {Barmby}, {Baumbach}, {Berry}, {Biscani}, {Boquien}, {Bostroem}, {Bouma}, {Brammer}, {Bray}, {Breytenbach}, {Buddelmeijer}, {Burke}, {Calderone}, {Cano Rodr{\'\i}guez}, {Cara}, {Cardoso}, {Cheedella}, {Copin}, {Corrales}, {Crichton}, {D'Avella}, {Deil}, {Depagne}, {Dietrich}, {Donath}, {Droettboom}, {Earl}, {Erben}, {Fabbro}, {Ferreira}, {Finethy}, {Fox}, {Garrison}, {Gibbons}, {Goldstein}, {Gommers}, {Greco}, {Greenfield}, {Groener}, {Grollier}, {Hagen}, {Hirst}, {Homeier}, {Horton}, {Hosseinzadeh}, {Hu}, {Hunkeler}, {Ivezi{\'c}}, {Jain}, {Jenness}, {Kanarek}, {Kendrew}, {Kern}, {Kerzendorf}, {Khvalko}, {King}, {Kirkby}, {Kulkarni},
  {Kumar}, {Lee}, {Lenz}, {Littlefair}, {Ma}, {Macleod}, {Mastropietro}, {McCully}, {Montagnac}, {Morris}, {Mueller}, {Mumford}, {Muna}, {Murphy}, {Nelson}, {Nguyen}, {Ninan}, {N{\"o}the}, {Ogaz}, {Oh}, {Parejko}, {Parley}, {Pascual}, {Patil}, {Patil}, {Plunkett}, {Prochaska}, {Rastogi}, {Reddy Janga}, {Sabater}, {Sakurikar}, {Seifert}, {Sherbert}, {Sherwood-Taylor}, {Shih}, {Sick}, {Silbiger}, {Singanamalla}, {Singer}, {Sladen}, {Sooley}, {Sornarajah}, {Streicher}, {Teuben}, {Thomas}, {Tremblay}, {Turner}, {Terr{\'o}n}, {van Kerkwijk}, {de la Vega}, {Watkins}, {Weaver}, {Whitmore}, {Woillez}, {Zabalza}, \& {Astropy Contributors}}]{astropy2018}
{Astropy Collaboration}, {Price-Whelan}, A.~M., {Sip{\H{o}}cz}, B.~M., {et~al.} 2018, \aj, 156, 123, \dodoi{10.3847/1538-3881/aabc4f}

\bibitem[{{B{\"o}hringer} {et~al.}(2002){B{\"o}hringer}, {Matsushita}, {Churazov}, {Ikebe}, \& {Chen}}]{Bohringer2002A&A}
{B{\"o}hringer}, H., {Matsushita}, K., {Churazov}, E., {Ikebe}, Y., \& {Chen}, Y. 2002, \aap, 382, 804, \dodoi{10.1051/0004-6361:20011708}

\bibitem[{{Bondi}(1952)}]{Bondi1952MNRAS}
{Bondi}, H. 1952, \mnras, 112, 195, \dodoi{10.1093/mnras/112.2.195}

\bibitem[{{Borgani} {et~al.}(2002){Borgani}, {Governato}, {Wadsley}, {Menci}, {Tozzi}, {Quinn}, {Stadel}, \& {Lake}}]{Borgani2002MNRAS}
{Borgani}, S., {Governato}, F., {Wadsley}, J., {et~al.} 2002, \mnras, 336, 409, \dodoi{10.1046/j.1365-8711.2002.05746.x}

\bibitem[{{Cant{\'o}} {et~al.}(2000){Cant{\'o}}, {Raga}, \& {Rodr{\'\i}guez}}]{Canto2000}
{Cant{\'o}}, J., {Raga}, A.~C., \& {Rodr{\'\i}guez}, L.~F. 2000, \apj, 536, 896, \dodoi{10.1086/308983}

\bibitem[{{Chevalier} \& {Clegg}(1985)}]{Chevalier1985}
{Chevalier}, R.~A., \& {Clegg}, A.~W. 1985, \nat, 317, 44, \dodoi{10.1038/317044a0}

\bibitem[{{Cho} {et~al.}(2023){Cho}, {Prather}, {Narayan}, {Natarajan}, {Su}, {Ricarte}, \& {Chatterjee}}]{Cho2023}
{Cho}, H., {Prather}, B.~S., {Narayan}, R., {et~al.} 2023, \apjl, 959, L22, \dodoi{10.3847/2041-8213/ad1048}

\bibitem[{{Ciotti} {et~al.}(1991){Ciotti}, {D'Ercole}, {Pellegrini}, \& {Renzini}}]{Ciotti1991ApJ}
{Ciotti}, L., {D'Ercole}, A., {Pellegrini}, S., \& {Renzini}, A. 1991, \apj, 376, 380, \dodoi{10.1086/170289}

\bibitem[{{Ciotti} \& {Ostriker}(1997)}]{Ciotti1997ApJ}
{Ciotti}, L., \& {Ostriker}, J.~P. 1997, \apjl, 487, L105, \dodoi{10.1086/310902}

\bibitem[{{Ciotti} \& {Ostriker}(2001)}]{Ciotti2001ApJ}
---. 2001, \apj, 551, 131, \dodoi{10.1086/320053}

\bibitem[{Collette(2013)}]{collette_python_hdf5_2014}
Collette, A. 2013, Python and HDF5 (O'Reilly)

\bibitem[{{Conroy} {et~al.}(2009){Conroy}, {Gunn}, \& {White}}]{Conroy2009ApJ}
{Conroy}, C., {Gunn}, J.~E., \& {White}, M. 2009, \apj, 699, 486, \dodoi{10.1088/0004-637X/699/1/486}

\bibitem[{{Conroy} {et~al.}(2015){Conroy}, {van Dokkum}, \& {Kravtsov}}]{Conroy2015ApJ}
{Conroy}, C., {van Dokkum}, P.~G., \& {Kravtsov}, A. 2015, \apj, 803, 77, \dodoi{10.1088/0004-637X/803/2/77}

\bibitem[{{Crain} {et~al.}(2015){Crain}, {Schaye}, {Bower}, {Furlong}, {Schaller}, {Theuns}, {Dalla Vecchia}, {Frenk}, {McCarthy}, {Helly}, {Jenkins}, {Rosas-Guevara}, {White}, \& {Trayford}}]{Crain2015MNRAS}
{Crain}, R.~A., {Schaye}, J., {Bower}, R.~G., {et~al.} 2015, \mnras, 450, 1937, \dodoi{10.1093/mnras/stv725}

\bibitem[{{Dalla Vecchia} \& {Schaye}(2012)}]{Vecchia2012MNRAS}
{Dalla Vecchia}, C., \& {Schaye}, J. 2012, \mnras, 426, 140, \dodoi{10.1111/j.1365-2966.2012.21704.x}

\bibitem[{{Donahue} {et~al.}(2011){Donahue}, {de Messi{\`e}res}, {O'Connell}, {Voit}, {Hoffer}, {McNamara}, \& {Nulsen}}]{Donahue2011ApJ}
{Donahue}, M., {de Messi{\`e}res}, G.~E., {O'Connell}, R.~W., {et~al.} 2011, \apj, 732, 40, \dodoi{10.1088/0004-637X/732/1/40}

\bibitem[{{Draine}(2011)}]{Draine2011piimbook}
{Draine}, B.~T. 2011, {Physics of the Interstellar and Intergalactic Medium} (Princeton University Press)

\bibitem[{{Dunn} {et~al.}(2010){Dunn}, {Allen}, {Taylor}, {Shurkin}, {Gentile}, {Fabian}, \& {Reynolds}}]{Dunn2010MNRAS}
{Dunn}, R.~J.~H., {Allen}, S.~W., {Taylor}, G.~B., {et~al.} 2010, \mnras, 404, 180, \dodoi{10.1111/j.1365-2966.2010.16314.x}

\bibitem[{Fabian(1994)}]{fabian1994cooling}
Fabian, A.~C. 1994, Annual Review of Astronomy and Astrophysics, 32, 277

\bibitem[{{Fabian}(2012)}]{Fabian2012ARA&A}
{Fabian}, A.~C. 2012, \araa, 50, 455, \dodoi{10.1146/annurev-astro-081811-125521}

\bibitem[{{Gatuzz} {et~al.}(2023{\natexlab{a}}){Gatuzz}, {Sanders}, {Dennerl}, {Liu}, {Fabian}, {Pinto}, {Eckert}, {Russell}, {Tamura}, {Walker}, \& {ZuHone}}]{Gatuzz2023MNRASa}
{Gatuzz}, E., {Sanders}, J.~S., {Dennerl}, K., {et~al.} 2023{\natexlab{a}}, \mnras, 520, 4793, \dodoi{10.1093/mnras/stad447}

\bibitem[{{Gatuzz} {et~al.}(2023{\natexlab{b}}){Gatuzz}, {Sanders}, {Dennerl}, {Liu}, {Fabian}, {Pinto}, {Eckert}, {Walker}, \& {ZuHone}}]{Gatuzz2023MNRASb}
---. 2023{\natexlab{b}}, \mnras, 525, 6394, \dodoi{10.1093/mnras/stad2716}

\bibitem[{{Generozov} {et~al.}(2015){Generozov}, {Stone}, \& {Metzger}}]{Generozov2015MNRAS}
{Generozov}, A., {Stone}, N.~C., \& {Metzger}, B.~D. 2015, \mnras, 453, 775, \dodoi{10.1093/mnras/stv1607}

\bibitem[{{Grattafiori} {et~al.}(2024){Grattafiori}, {Dubey}, {Jauhri}, {Pandey}, {Kadian}, {Al-Dahle}, {Letman}, {Mathur}, {Schelten}, {Vaughan}, {Yang}, {Fan}, {Goyal}, {Hartshorn}, {Yang}, {Mitra}, {Sravankumar}, {Korenev}, {Hinsvark}, {Rao}, {Zhang}, {Rodriguez}, {Gregerson}, {Spataru}, {Roziere}, {Biron}, {Tang}, {Chern}, {Caucheteux}, {Nayak}, {Bi}, {Marra}, {McConnell}, {Keller}, {Touret}, {Wu}, {Wong}, {Canton Ferrer}, {Nikolaidis}, {Allonsius}, {Song}, {Pintz}, {Livshits}, {Wyatt}, {Esiobu}, {Choudhary}, {Mahajan}, {Garcia-Olano}, {Perino}, {Hupkes}, {Lakomkin}, {AlBadawy}, {Lobanova}, {Dinan}, {Smith}, {Radenovic}, {Guzm{\'a}n}, {Zhang}, {Synnaeve}, {Lee}, {Anderson}, {Thattai}, {Nail}, {Mialon}, {Pang}, {Cucurell}, {Nguyen}, {Korevaar}, {Xu}, {Touvron}, {Zarov}, {Arrieta Ibarra}, {Kloumann}, {Misra}, {Evtimov}, {Zhang}, {Copet}, {Lee}, {Geffert}, {Vranes}, {Park}, {Mahadeokar}, {Shah}, {van der Linde}, {Billock}, {Hong}, {Lee}, {Fu}, {Chi}, {Huang}, {Liu}, {Wang}, {Yu}, {Bitton}, {Spisak}, {Park},
  {Rocca}, {Johnstun}, {Saxe}, {Jia}, {Vasuden Alwala}, {Prasad}, {Upasani}, {Plawiak}, {Li}, {Heafield}, {Stone}, {El-Arini}, {Iyer}, {Malik}, {Chiu}, {Bhalla}, {Lakhotia}, {Rantala-Yeary}, {van der Maaten}, {Chen}, {Tan}, {Jenkins}, {Martin}, {Madaan}, {Malo}, {Blecher}, {Landzaat}, {de Oliveira}, {Muzzi}, {Pasupuleti}, {Singh}, {Paluri}, {Kardas}, {Tsimpoukelli}, {Oldham}, {Rita}, {Pavlova}, {Kambadur}, {Lewis}, {Si}, {Singh}, {Hassan}, {Goyal}, {Torabi}, {Bashlykov}, {Bogoychev}, {Chatterji}, {Zhang}, {Duchenne}, {{\c{C}}elebi}, {Alrassy}, {Zhang}, {Li}, {Vasic}, {Weng}, {Bhargava}, {Dubal}, {Krishnan}, {Singh Koura}, {Xu}, {He}, {Dong}, {Srinivasan}, {Ganapathy}, {Calderer}, {Silveira Cabral}, {Stojnic}, {Raileanu}, {Maheswari}, {Girdhar}, {Patel}, {Sauvestre}, {Polidoro}, {Sumbaly}, {Taylor}, {Silva}, {Hou}, {Wang}, {Hosseini}, {Chennabasappa}, {Singh}, {Bell}, {Kim}, {Edunov}, {Nie}, {Narang}, {Raparthy}, {Shen}, {Wan}, {Bhosale}, {Zhang}, {Vandenhende}, {Batra}, {Whitman}, {Sootla}, {Collot},
  {Gururangan}, {Borodinsky}, {Herman}, {Fowler}, {Sheasha}, {Georgiou}, {Scialom}, \& {Speckbacher}}]{LLamaMetaAI}
{Grattafiori}, A., {Dubey}, A., {Jauhri}, A., {et~al.} 2024, arXiv e-prints, arXiv:2407.21783, \dodoi{10.48550/arXiv.2407.21783}

\bibitem[{{Guo} {et~al.}(2023){Guo}, {Stone}, {Kim}, \& {Quataert}}]{MGuo2023ApJ}
{Guo}, M., {Stone}, J.~M., {Kim}, C.-G., \& {Quataert}, E. 2023, \apj, 946, 26, \dodoi{10.3847/1538-4357/acb81e}

\bibitem[{Harris {et~al.}(2020)Harris, Millman, van~der Walt, Gommers, Virtanen, Cournapeau, Wieser, Taylor, Berg, Smith, Kern, Picus, Hoyer, van Kerkwijk, Brett, Haldane, del R{\'{i}}o, Wiebe, Peterson, G{\'{e}}rard-Marchant, Sheppard, Reddy, Weckesser, Abbasi, Gohlke, \& Oliphant}]{Harris2020}
Harris, C.~R., Millman, K.~J., van~der Walt, S.~J., {et~al.} 2020, {Array programming with NumPy},  Nature Research, \dodoi{10.1038/s41586-020-2649-2}

\bibitem[{Hunter(2007)}]{Hunter4160265}
Hunter, J.~D. 2007, Computing in Science Engineering, 9, 90, \dodoi{10.1109/MCSE.2007.55}

\bibitem[{{Lemaster} \& {Stone}(2009)}]{Lemaster2009ApJ}
{Lemaster}, M.~N., \& {Stone}, J.~M. 2009, \apj, 691, 1092, \dodoi{10.1088/0004-637X/691/2/1092}

\bibitem[{{Li} {et~al.}(2020{\natexlab{a}}){Li}, {Li}, {Bryan}, {Ostriker}, \& {Quataert}}]{MLi2020ApJa}
{Li}, M., {Li}, Y., {Bryan}, G.~L., {Ostriker}, E.~C., \& {Quataert}, E. 2020{\natexlab{a}}, \apj, 894, 44, \dodoi{10.3847/1538-4357/ab86b4}

\bibitem[{{Li} {et~al.}(2020{\natexlab{b}}){Li}, {Li}, {Bryan}, {Ostriker}, \& {Quataert}}]{MLi2020ApJb}
---. 2020{\natexlab{b}}, \apj, 898, 23, \dodoi{10.3847/1538-4357/ab9c22}

\bibitem[{{Li} {et~al.}(2019){Li}, {Bryan}, \& {Quataert}}]{YLi2019ApJ}
{Li}, Y., {Bryan}, G.~L., \& {Quataert}, E. 2019, \apj, 887, 41, \dodoi{10.3847/1538-4357/ab4bca}

\bibitem[{{Maoz} \& {Graur}(2017)}]{Maoz2017ApJ}
{Maoz}, D., \& {Graur}, O. 2017, \apj, 848, 25, \dodoi{10.3847/1538-4357/aa8b6e}

\bibitem[{{Mathews} \& {Baker}(1971)}]{Mathews1971ApJ}
{Mathews}, W.~G., \& {Baker}, J.~C. 1971, \apj, 170, 241, \dodoi{10.1086/151208}

\bibitem[{{Mathews} \& {Loewenstein}(1986)}]{Mathews1986ApJ}
{Mathews}, W.~G., \& {Loewenstein}, M. 1986, \apjl, 306, L7, \dodoi{10.1086/184693}

\bibitem[{{McNamara} \& {Nulsen}(2007)}]{mcnamara2007}
{McNamara}, B.~R., \& {Nulsen}, P.~E.~J. 2007, \araa, 45, 117, \dodoi{10.1146/annurev.astro.45.051806.110625}

\bibitem[{{McNamara} \& {Nulsen}(2012)}]{McNamara2012NJPh}
---. 2012, New Journal of Physics, 14, 055023, \dodoi{10.1088/1367-2630/14/5/055023}

\bibitem[{{Merritt} \& {Ferrarese}(2001)}]{Merritt2001MNRAS}
{Merritt}, D., \& {Ferrarese}, L. 2001, \mnras, 320, L30, \dodoi{10.1046/j.1365-8711.2001.04165.x}

\bibitem[{{Mohapatra} \& {Quataert}(2024)}]{Mohapatra2024ApJ}
{Mohapatra}, R., \& {Quataert}, E. 2024, \apj, 965, 105, \dodoi{10.3847/1538-4357/ad2940}

\bibitem[{{Negri} {et~al.}(2014){Negri}, {Posacki}, {Pellegrini}, \& {Ciotti}}]{Negri2014MNRAS445}
{Negri}, A., {Posacki}, S., {Pellegrini}, S., \& {Ciotti}, L. 2014, \mnras, 445, 1351, \dodoi{10.1093/mnras/stu1834}

\bibitem[{Okuta {et~al.}(2017)Okuta, Unno, Nishino, Hido, \& Loomis}]{Okuta2017CuPyA}
Okuta, R., Unno, Y., Nishino, D., Hido, S., \& Loomis, C. 2017, in Proceedings of Workshop on Machine Learning Systems (LearningSys) in The Thirty-first Annual Conference on Neural Information Processing Systems (NIPS).
\newblock \url{http://learningsys.org/nips17/assets/papers/paper_16.pdf}

\bibitem[{{Pillepich} {et~al.}(2018){Pillepich}, {Springel}, {Nelson}, {Genel}, {Naiman}, {Pakmor}, {Hernquist}, {Torrey}, {Vogelsberger}, {Weinberger}, \& {Marinacci}}]{Pillepich2018MNRAS}
{Pillepich}, A., {Springel}, V., {Nelson}, D., {et~al.} 2018, \mnras, 473, 4077, \dodoi{10.1093/mnras/stx2656}

\bibitem[{{Richtler} {et~al.}(2008){Richtler}, {Schuberth}, {Hilker}, {Dirsch}, {Bassino}, \& {Romanowsky}}]{Richtler2008A&A}
{Richtler}, T., {Schuberth}, Y., {Hilker}, M., {et~al.} 2008, \aap, 478, L23, \dodoi{10.1051/0004-6361:20078539}

\bibitem[{{Schaye} {et~al.}(2015){Schaye}, {Crain}, {Bower}, {Furlong}, {Schaller}, {Theuns}, {Dalla Vecchia}, {Frenk}, {McCarthy}, {Helly}, {Jenkins}, {Rosas-Guevara}, {White}, {Baes}, {Booth}, {Camps}, {Navarro}, {Qu}, {Rahmati}, {Sawala}, {Thomas}, \& {Trayford}}]{Schaye2015MNRAS}
{Schaye}, J., {Crain}, R.~A., {Bower}, R.~G., {et~al.} 2015, \mnras, 446, 521, \dodoi{10.1093/mnras/stu2058}

\bibitem[{{Schuberth} {et~al.}(2010){Schuberth}, {Richtler}, {Hilker}, {Dirsch}, {Bassino}, {Romanowsky}, \& {Infante}}]{Schuberth2010A&A}
{Schuberth}, Y., {Richtler}, T., {Hilker}, M., {et~al.} 2010, \aap, 513, A52, \dodoi{10.1051/0004-6361/200912482}

\bibitem[{{Schure} {et~al.}(2009){Schure}, {Kosenko}, {Kaastra}, {Keppens}, \& {Vink}}]{Schure2009A&A}
{Schure}, K.~M., {Kosenko}, D., {Kaastra}, J.~S., {Keppens}, R., \& {Vink}, J. 2009, \aap, 508, 751, \dodoi{10.1051/0004-6361/200912495}

\bibitem[{{Shen} {et~al.}(2003){Shen}, {Mo}, {White}, {Blanton}, {Kauffmann}, {Voges}, {Brinkmann}, \& {Csabai}}]{Shen2003}
{Shen}, S., {Mo}, H.~J., {White}, S. D.~M., {et~al.} 2003, \mnras, 343, 978, \dodoi{10.1046/j.1365-8711.2003.06740.x}

\bibitem[{{Stone} {et~al.}(2020){Stone}, {Tomida}, {White}, \& {Felker}}]{Stone2020ApJS}
{Stone}, J.~M., {Tomida}, K., {White}, C.~J., \& {Felker}, K.~G. 2020, \apjs, 249, 4, \dodoi{10.3847/1538-4365/ab929b}

\bibitem[{{Stone} {et~al.}(2024){Stone}, {Mullen}, {Fielding}, {Grete}, {Guo}, {Kempski}, {Most}, {White}, \& {Wong}}]{Stone2024arXiv}
{Stone}, J.~M., {Mullen}, P.~D., {Fielding}, D., {et~al.} 2024, arXiv e-prints, arXiv:2409.16053, \dodoi{10.48550/arXiv.2409.16053}

\bibitem[{{Su} {et~al.}(2017){Su}, {Nulsen}, {Kraft}, {Forman}, {Jones}, {Irwin}, {Randall}, \& {Churazov}}]{Su2017ApJ}
{Su}, Y., {Nulsen}, P. E.~J., {Kraft}, R.~P., {et~al.} 2017, \apj, 847, 94, \dodoi{10.3847/1538-4357/aa8954}

\bibitem[{{Tang} {et~al.}(2009{\natexlab{a}}){Tang}, {Wang}, {Lu}, \& {Mo}}]{Tang2009MNRASa}
{Tang}, S., {Wang}, Q.~D., {Lu}, Y., \& {Mo}, H.~J. 2009{\natexlab{a}}, \mnras, 392, 77, \dodoi{10.1111/j.1365-2966.2008.14057.x}

\bibitem[{{Tang} {et~al.}(2009{\natexlab{b}}){Tang}, {Wang}, {Mac Low}, \& {Joung}}]{Tang2009MNRASb}
{Tang}, S., {Wang}, Q.~D., {Mac Low}, M.-M., \& {Joung}, M.~R. 2009{\natexlab{b}}, \mnras, 398, 1468, \dodoi{10.1111/j.1365-2966.2009.15206.x}

\bibitem[{{Trott} {et~al.}(2021){Trott}, {Berger-Vergiat}, {Poliakoff}, {Rajamanickam}, {Lebrun-Grandie}, {Madsen}, {Al Awar}, {Gligoric}, {Shipman}, \& {Womeldorff}}]{Trott2021CSE}
{Trott}, C., {Berger-Vergiat}, L., {Poliakoff}, D., {et~al.} 2021, Computing in Science and Engineering, 23, 10, \dodoi{10.1109/MCSE.2021.3098509}

\bibitem[{{van der Velden}(2020)}]{Ellert2020JOSS}
{van der Velden}, E. 2020, The Journal of Open Source Software, 5, 2004, \dodoi{10.21105/joss.02004}

\bibitem[{Virtanen {et~al.}(2020)Virtanen, Gommers, Oliphant, Haberland, Reddy, Cournapeau, Burovski, Peterson, Weckesser, Bright, van~der Walt, Brett, Wilson, Millman, Mayorov, Nelson, Jones, Kern, Larson, Carey, Polat, Feng, Moore, VanderPlas, Laxalde, Perktold, Cimrman, Henriksen, Quintero, Harris, Archibald, Ribeiro, Pedregosa, van Mulbregt, Vijaykumar, Bardelli, Rothberg, Hilboll, Kloeckner, Scopatz, Lee, Rokem, Woods, Fulton, Masson, H{\"a}ggstr{\"o}m, Fitzgerald, Nicholson, Hagen, Pasechnik, Olivetti, Martin, Wieser, Silva, Lenders, Wilhelm, Young, Price, Ingold, Allen, Lee, Audren, Probst, Dietrich, Silterra, Webber, Slavi{\v c}, Nothman, Buchner, Kulick, Sch{\"o}nberger, de~Miranda~Cardoso, Reimer, Harrington, Rodr{\'\i}guez, Nunez-Iglesias, Kuczynski, Tritz, Thoma, Newville, K{\"u}mmerer, Bolingbroke, Tartre, Pak, Smith, Nowaczyk, Shebanov, Pavlyk, Brodtkorb, Lee, McGibbon, Feldbauer, Lewis, Tygier, Sievert, Vigna, Peterson, More, Pudlik, Oshima, Pingel, Robitaille, Spura, Jones, Cera, Leslie, Zito,
  Krauss, Upadhyay, Halchenko, V{\'a}zquez-Baeza, \& Contributors}]{Virtanen2020}
Virtanen, P., Gommers, R., Oliphant, T.~E., {et~al.} 2020, Nature Methods, 17, 261, \dodoi{10.1038/s41592-019-0686-2}

\bibitem[{{Voit} {et~al.}(2015){Voit}, {Donahue}, {O'Shea}, {Bryan}, {Sun}, \& {Werner}}]{Voit2015ApJ803L21V}
{Voit}, G.~M., {Donahue}, M., {O'Shea}, B.~W., {et~al.} 2015, \apjl, 803, L21, \dodoi{10.1088/2041-8205/803/2/L21}

\bibitem[{{Voit} {et~al.}(2020){Voit}, {Bryan}, {Prasad}, {Frisbie}, {Li}, {Donahue}, {O'Shea}, {Sun}, \& {Werner}}]{Voit2020ApJ}
{Voit}, G.~M., {Bryan}, G.~L., {Prasad}, D., {et~al.} 2020, \apj, 899, 70, \dodoi{10.3847/1538-4357/aba42e}

\bibitem[{{Wang} {et~al.}(2019){Wang}, {Li}, \& {Ruszkowski}}]{CWang2019MNRAS}
{Wang}, C., {Li}, Y., \& {Ruszkowski}, M. 2019, \mnras, 482, 3576, \dodoi{10.1093/mnras/sty2906}

\bibitem[{{Werner} {et~al.}(2012){Werner}, {Allen}, \& {Simionescu}}]{Werner2012MNRAS}
{Werner}, N., {Allen}, S.~W., \& {Simionescu}, A. 2012, \mnras, 425, 2731, \dodoi{10.1111/j.1365-2966.2012.21245.x}

\end{thebibliography}
\end{CJK*}
\end{document}